\documentclass[aps,pra,twocolumn]{revtex4-2}
\usepackage{graphicx,color,amsmath,amssymb}

\usepackage{xcolor}
\usepackage[export]{adjustbox}

\begin{document}
\title{Optimizing persistent currents in a ring-shaped Bose-Einstein condensate using machine learning}
\author{Simeon~Simjanovski}
\affiliation{ARC Centre of Excellence for Engineered Quantum Systems, School of Mathematics and Physics, University of Queensland, St Lucia, QLD 4072, Australia}

\author{Guillaume~Gauthier}
\affiliation{ARC Centre of Excellence for Engineered Quantum Systems, School of Mathematics and Physics, University of Queensland, St Lucia, QLD 4072, Australia}

\author{Matthew~J.~Davis}
\affiliation{ARC Centre of Excellence for Engineered Quantum Systems, School of Mathematics and Physics, University of Queensland, St Lucia, QLD 4072, Australia}

\author{Halina~Rubinsztein-Dunlop}
\affiliation{ARC Centre of Excellence for Engineered Quantum Systems, School of Mathematics and Physics, University of Queensland, St Lucia, QLD 4072, Australia}

\author{Tyler~W.~Neely}
\affiliation{ARC Centre of Excellence for Engineered Quantum Systems, School of Mathematics and Physics, University of Queensland, St Lucia, QLD 4072, Australia}
\date{\today{}}

\begin{abstract}
We demonstrate a method for generating persistent currents in Bose-Einstein condensates by using a Gaussian process learner to experimentally control the stirring of the superfluid. The learner optimizes four different outcomes of the stirring process: (O.I) targeting and (O.II) maximization of the persistent current winding number; and (O.III) targeting and (O.IV) maximization with time constraints. The learner optimizations are determined based on the achieved winding number and the number of spurious vortices introduced by stirring. We find that the learner is successful in optimizing the stirring protocols, although the optimal stirring profiles vary significantly depending strongly on the choice of cost function and scenario. These results suggest that stirring is robust and persistent currents can be reliably generated through a variety of stirring approaches.
\end{abstract}

\maketitle

\section{\label{sec:background}Introduction and Background}
In a Bose-Einstein condensate (BEC) superfluid, a persistent current (PC) is the quantized circulation of the bulk fluid in a multiply connected geometry,  such as a ring~\cite{leggett2001bose,ryu2007observation}. As the phase singularity associated with the PC is located at the center of the ring, where there is no density, the flow is stabilized by topology against dissipation~\cite{leggett2001bose,neely2013characteristics}. While PCs are of fundamental interest in BEC research, they are also relevant to atomtronics~\cite{amico2021roadmap,Amico_2022}, where they can form the basis of compact matter-wave interferometers~\cite{Ryu_2013,halkyard2010rotational,Helm_2018}. PCs are also the basis of fluxon systems~\cite{Oliinyk_2019} that have potential applications in atomtronics~\cite{Oliinyk_2020}, for example as qubits which can be physically relocated~\cite{Kaurov_2005}.
 
Multiple methods for experimentally generating PCs have been developed. One can imprint the desired currents directly to the wave-function phase using an optical potential~\cite{Denschlag_2000,del2022imprinting}, inducing superfluid flow proportional to the phase gradient~\cite{leggett2001bose}. However, in doing so the phase profile of the condensate acquires a sharp $2N\pi$ phase jump for a charge-$N$ current at some point along the ring leading to significant density excitations~\cite{del2022imprinting,Kumar_2018}. Alternatively, orbital angular momentum can be directly transferred from Laguerre-Gaussian laser modes~\cite{Allen_1992,beattie2013persistent}, to the condensate, although the typical efficiency is limited by mode matching to around 50\%~\cite{Andersen_2006}. A more straightforward method is to stir the condensate using an optical barrier, analogous to stirring a classical fluid to generate flow~\cite{wright2013driving,Eckel_2014_Hysteresis}. In this paper, we explore the parameter space of a stirring method, with an aim to determine optimal stirring profiles with a machine learning approach. Determining the best-case stirring scenarios will highlight the advantages and disadvantages of this approach compared to the other schemes.

Machine learning has increasingly been used for the experimental control of quantum systems~\cite{Carleo_2019,Zhou_2018,Barker_2020}. The key advantage of this approach is the lack of prior knowledge a machine learner has on the system of interest, allowing for unique or counter-intuitive solutions to a given problem to be found~\cite{Zhou_2018,Barker_2020,MendelsonShahar2003ALoM,Jones_1998}. These solutions can be found even when there is an incomplete physical model of the system available, since the learner requires only a restricted set of data points~\cite{BonaccorsoGiuseppe2018MMLA}. Furthermore, computers have the ability to handle large amounts of data efficiently as compared to human operators. These factors have established machine learning as a potentially superior method of experimental control~\cite{Nakamura:19}. 
\begin{figure*}[t!]
    \centering
    \includegraphics[trim={2cm 0cm 2cm 0cm},clip,width=\textwidth]{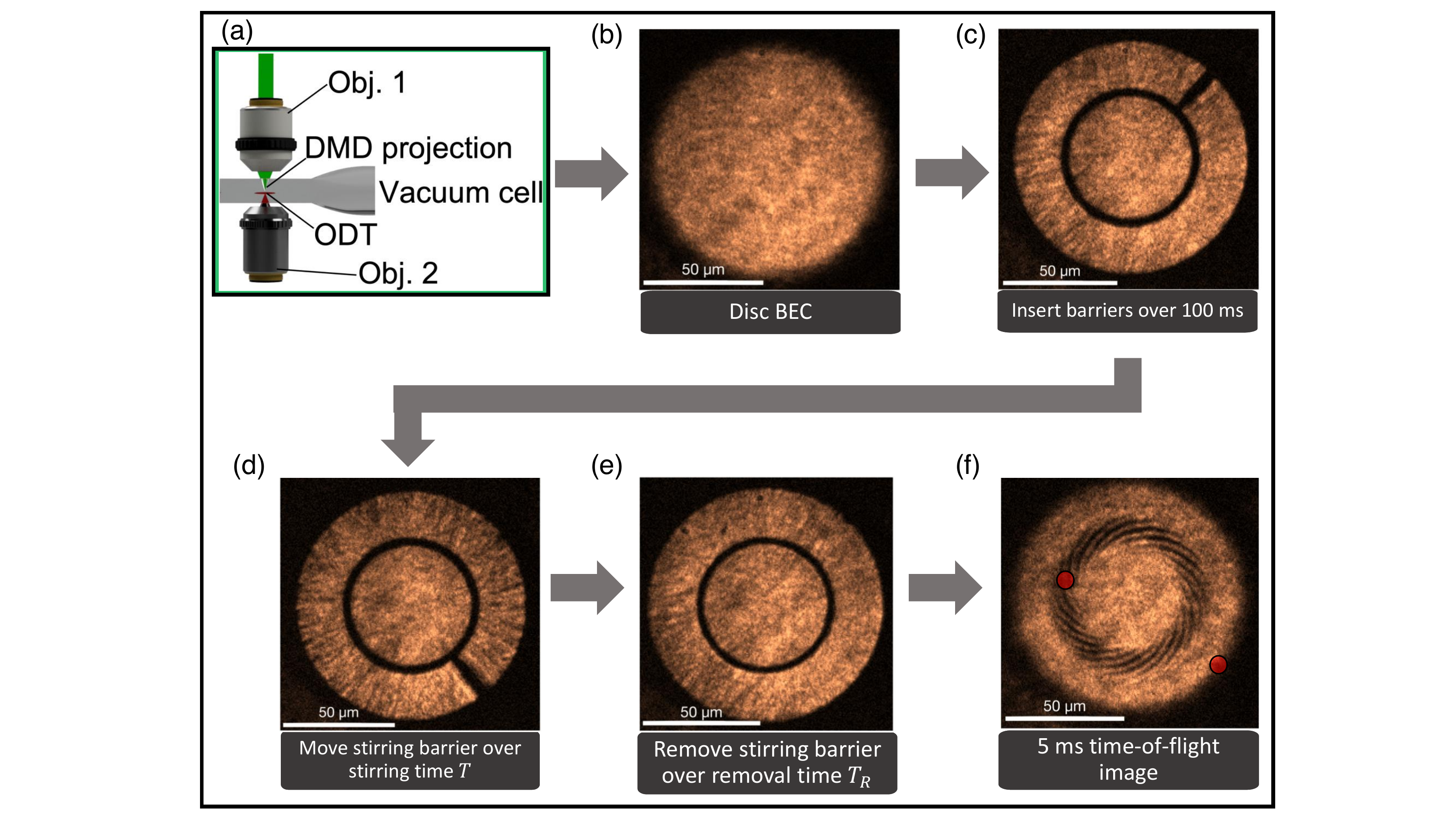}
    \caption {(a) A simplified diagram of the main apparatus used to generate and control BECs. The light from the digital micromirror device (DMD) is shown by the green beam along the projection (Obj. 1) and imaging (Obj. 2) objectives. The light used for Faraday imaging propagates along the same direction. (b–e) Experimental images demonstrating the stirring process using the DMD projected light. The ring trap shown here has a larger width than the subsequent optimizations for illustrative clarity. (f) A typical time of flight experiment using stirring as described in Sec. II, with red dots indicating spurious vortices.}
    \label{fig:Setup}
\end{figure*}

The stirring of a BEC to generate a PC is complicated by a variety of experimental realities that can be challenging or inefficient to model. For example, there may be density variations due to trap imperfections or roughness that can cause the PC to degrade or decay~\cite{Bell_2016}. The BEC also has a finite lifetime within which stirring, system dynamics, and data extraction must occur, introducing time restrictions to the stirring process~\cite{Gauthier}. Aggressive stirring of the system will likely result in the generation of undesirable excitations such as quantum vortices and phonons~\cite{Wright_2013}. These aspects suggest the stirring can be a complex process covering a large parameter space that may be challenging to accurately describe and control. Thus, under a set of desired  constraints, optimal stirring parameters for this complex system are not \textit{a priori} obvious. For example, various approaches have moved the barrier at constant speed~\cite{Wright_2013,Yakimenko_2015_VE} or accelerating~\cite{Murray_2013} speed profiles, but it is unclear which approach is superior. The stirring barrier must also be removed after inducing flow at some optimal rate. Finally, one might be interested in creating the current in the shortest time possible, to minimize the effect of atom loss due to a finite BEC lifetime. This large parameter space motivates the use of machine learning approaches to optimize the stirring process, subject to user-chosen sets of constraints. Recently, machine learning optimization of persistent currents in a ring lattice has been explored in numerical simulations~\cite{Amico_2021}, but such approaches have not yet been applied to experiments. 

Here, we use reinforcement learning through the open-source Machine-Learning Online Optimization Package (M-LOOP)~\cite{Wigley_2016} to explore the stirring parameter space. Using the Gaussian process learner in M-LOOP and applying it to four separate optimization conditions, we find that the learner is able to successfully optimize the stirring process, generating the desired PCs while minimizing spurious vortices. However, we find that the optimized parameters vary widely depending on the chosen cost function and target parameters. The results suggest that stirring is robust against variations in the particular details of the stirring profile. 
\section{\label{sec:BECCreation}Experimental Implementation}
\begin{figure*}[t!]
    \centering
    \includegraphics[trim={0cm 3.6cm 0cm  1.15cm},clip,width=0.9\textwidth]{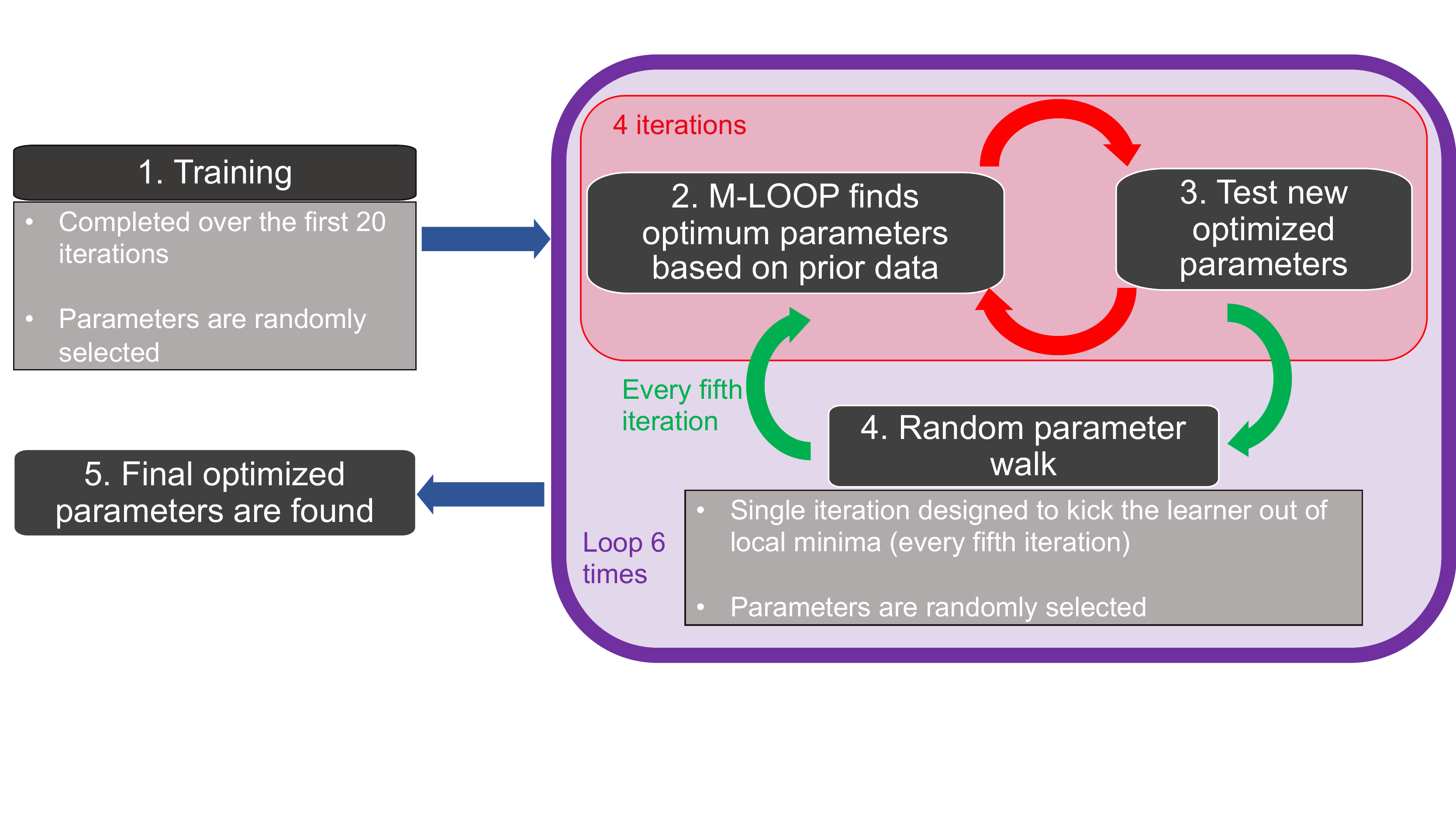}
    \caption{Process flow chart showing the general stirring parameter optimization process using M-LOOP over 50 iterations. Each iteration is in reality composed of five experimental runs, taken with the same parameter settings. The cost of each iteration is computed as the average cost for these five runs. This results in a mean cost and uncertainty (standard error) returned to M-LOOP for each parameter setting.}
    \label{fig:ProcessFlow}
\end{figure*}
Our experiment implements configured optical potentials with blue-detuned repulsive light to both trap a $^{87}$Rb BEC in a ring and simultaneously stir it, as shown in Fig.~\ref{fig:Setup}. The optical potential for the ring geometry and stirring is generated using a digital micromirror device (DMD). The DMD is an array composed of 2560 $\times$ 1600 mirrors that can be individually switched ``on'' and ``off'' corresponding to $\pm 12^{\circ}$ tilts, respectively. This enables the creation of a ``binary" hard-wall trapping potential~\cite{Gauthier:16}. A detailed description of the process for creating and configuring BECs using this apparatus can be found in our prior work~\cite{Gauthier_2020,Gauthier:16}. 

The experimental sequence is shown in Fig.~\ref{fig:Setup}. The atoms are initially trapped in a $100$-$\mu$m diameter disk. A separation barrier is then inserted, growing from zero to $7$-$\mu$m width over $100$-ms symmetrically about a mean radius of $34$-$\mu$m. This insertion method was observed to be sufficiently slow to avoid density excitations. This process results in two separated condensates: an outer ring BEC where the PC is created, and an inner reference BEC that has a uniform phase, and can be interfered with the outer BEC resulting in a pattern from which the PC winding number can be inferred~\cite{Andrews_1997}. Simultaneously, a $3$-$\mu$m thick stirring barrier is inserted into the ring BEC by linearly increasing the barrier thickness from $0$-$\mu$m. This barrier is used for the stirring of the outer ring, subject to machine learning controlled parameters (see Section~\ref{sec:MLConcept}) which aim to generate the desired persistent current. After the stirring process is completed, the stirring barrier is removed by linearly decreasing its thickness over the removal time $T_R$ while maintaining the same angular velocity. The stirring region and reference BEC are then interfered by releasing the system from the trap, and imaged in time of flight (TOF)~\cite{Eckel_2014,Eckel_2014_Hysteresis,del2022imprinting}.

The number of spiral fringes resulting from the interference of the ring and reference condensates identifies the phase winding number of the ring, which is directly proportional to the PC flow speed~\cite{Eckel_2014}. Spuriously generated quantum vortices can form due to the stirring and are resolvable as density dips in TOF, as seen in Fig.~\ref{fig:Setup}. A single TOF image can thus be used to determine both the winding and vortex number. A detailed description of  the image analysis is given in Appendix~\ref{sec:Imaging}.

\section{\label{sec:MLConcept}Machine Learning Implementation}
\subsection{Stirrer parametrization}
Machine learning control of the stirring is implemented with Gaussian process regression via the freely available package M-LOOP~\cite{Wigley_2016,Wigley_Thesis}. As a reinforcement machine learning method, M-LOOP requires an external measure of the error produced by the parameter model chosen by the learner \cite{BonaccorsoGiuseppe2018MMLA,MEHTA20191}. This error can be quantified using a cost function. The choice of cost function can vary specific to the type of optimization required, but there is one underlying property across all choices: the cost should be minimized when the parameter set is optimized \cite{Jones_1998,Carleo_2019}.

We begin by specifying the angular stirring function of the barrier:
\begin{align}
    \theta(t)=\alpha \left(\frac{t}{A}\right)^P, \label{eqn:StirProf1}
\end{align}
where the constant $A = 400$-ms is introduced to ensure the stirring is slow enough to prevent destroying the BEC. The parameters controlled by the machine learner are as follows: (1) \textit{Stirring time} of the barrier after insertion, $T$; (2) \textit{Removal time} of the barrier after the stirring time, $T_R$; (3) \textit{Stirring exponent}, $P$; and (4) \textit{Angular coefficient}, $\alpha$. Each parameter is given a lower and upper bound: (1) $100$ ms $\leq T\leq 1500$ ms, (2) $10$ ms $\leq T_R\leq450$ ms, (3) $1 \leq P \leq 4.5$ [dimensionless], and (4) $1.75\times10^{-2}$-rad $\leq \alpha\leq\alpha_\textrm{max}$, where $\alpha_\textrm{max} = 3.49$-rad in the case of optimizations considered in Sec.~\ref{sec:MLConcept} and $\alpha_\textrm{max} = 10.5$-rad for those considered in Sec.~\ref{sec:MinimumTime}. This was done to restrict the parameter space initially (Sec.~\ref{sec:TUOptimisation}) for a simpler optimizations while increasing complexity with a larger parameter space in the later optimizations (Sec.~\ref{sec:MinimumTime}) to test the limitations of the learner employed. The lower bound on $T$ is selected so that even for the minimum stirring time sufficient fringes are generated for the detection algorithm to operate accurately~\footnote{With very few fringes present, the fringes themselves wrap around the entire interference region. The detection algorithm effectively counts periodicity of density in these regions meaning that such wrapping obscures the measurement.} (see also Appendix~\ref{sec:Imaging}). The upper bound is imposed due to the frame-rate constraints on the DMD, ensuring that the sequence projected onto the BEC is smooth when the stirring occurs, avoiding the stirring barrier moving around the ring in large, discrete jumps. Bounds on $T_R$ are also chosen such that the removal time is comparable to the insertion time. The ranges for $\alpha$ and $P$ were chosen by observation to ensure parameters that both were sufficient to consistently generate observable spiral fringes and did not result in aggressive stirring such that the BEC was destroyed.

\subsection{Cost functions}
\begin{figure}[t!]
    \centering
\includegraphics[trim={0cm 5.2cm 0cm
0cm},clip,width=\hsize]{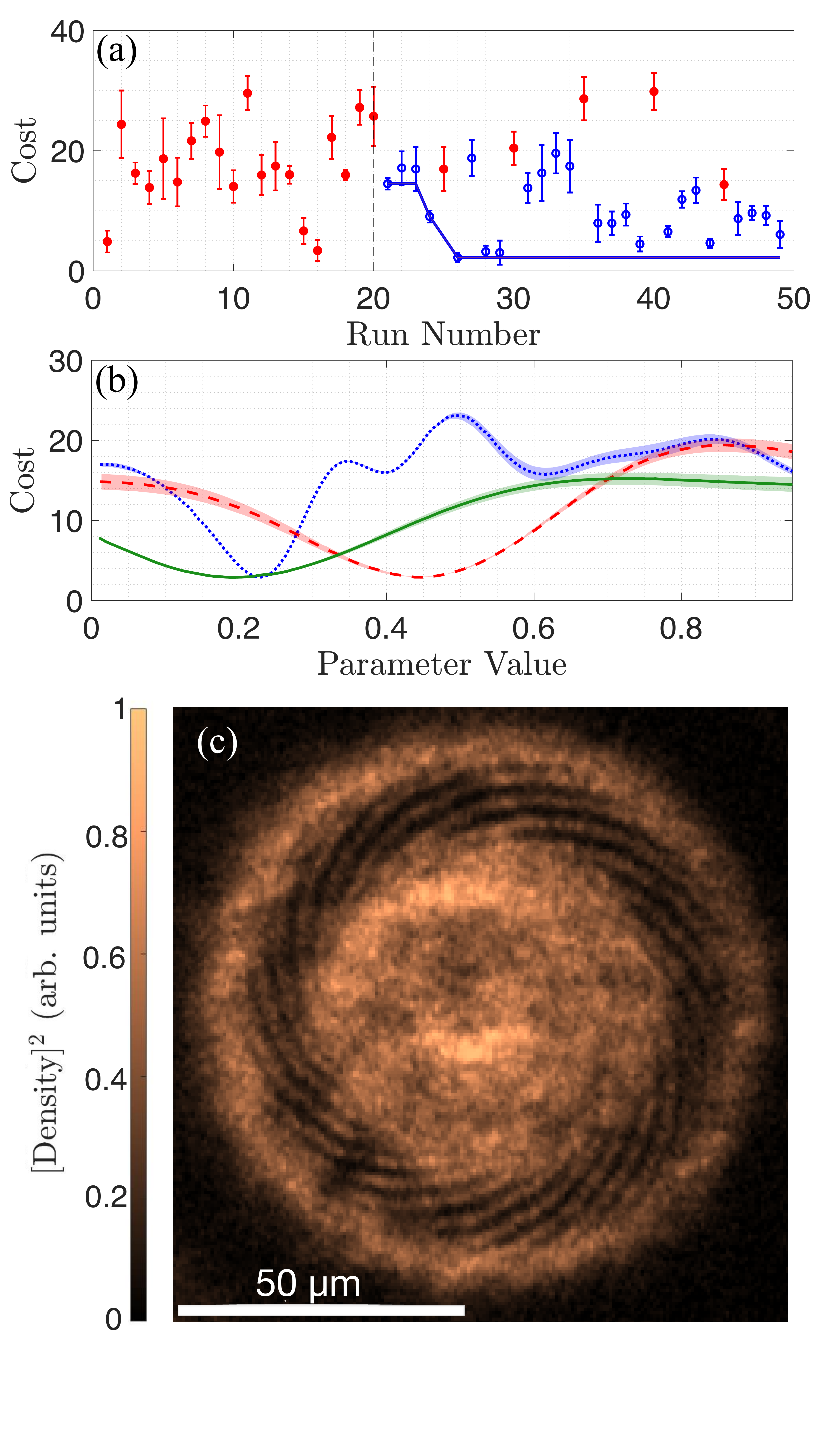}
\caption{M-LOOP output graphics for optimization to a target of $N_T=20$ windings (optimization O.I -- cost function given by Eq.~(\ref{eqn:CostTargeting})).  (a) Cost vs run number over 50 runs. The initial 20 random runs are shown before the vertical dashed line. Filled (red) points represent the randomly chosen parameter runs, and open (blue) points represent the runs with machine learner chosen parameters. Filled (red) points following the training can also be observed, which also represent random walks designed to drive the learner out of potential local minima while testing the optimum parameters. The solid (blue) line tracks the minimum cost achieved as a function of run number. (b) The predicted landscape for each of the three parameters computed using the costs.  The dashed (red) line represents the stirring time parameter $T$, the solid (green) line represents the removal time $T_R$, and the dotted (blue) line represents the angular coefficient $\alpha$. Parameter values are normalized to lower and upper ($P_{\text{min}}$ and $P_{\text{max}}$) bounds using $(P-P_{\text{min}})$/$(P_{\text{max}}-P_\text{min})$. Uncertainties, shown by the shaded regions for each curve, are computed through M-LOOP by feeding the learner an error estimate for each run. (c) The 5-ms TOF image resulting from using the optimum parameters in stirring. This particular image contains 19 fringes.}
\label{fig:TOData}
\end{figure}

Two main optimizations are considered:
targeting a specific winding number, with the cost function
\begin{align}
    \mathcal{C}_\textrm{T}&=\left|N_T-N_W\right|+N_V+10\frac{T}{T_{\text{max}}},\label{eqn:CostTargeting}
    \end{align}
    and maximizing the winding number, with the cost function
    \begin{align}
    \mathcal{C}_\textrm{M}&=-N_W+N_V+10\frac{T}{T_{\text{max}}}.\label{eqn:Costmaximization}
\end{align}
Here, the measured winding number $N_W$ and the measured spurious vortex number $N_V$ are obtained from the image analysis [Appendix~\ref{sec:Imaging}]. The target winding number $N_T$ is chosen prior to optimization. The maximum allowed stirring time $T_{\text{max}}$ is introduced for optimizations O.III and O.IV where we impose a time restriction for the stirring process (Sec.~\ref{sec:MinimumTime}). This time restriction on stirring is made so that any experiments which require both stirring and observation of subsequent dynamics of the BEC system can be made within the BEC lifetime, $\approx$~20~s for our system. By minimizing the stirring time, we make more time available for these subsequent dynamics to be observed. Otherwise, we set $T_{\text{max}}\to\infty$ such that the last term in the cost functions vanishes. Cost is computed using an average measure of cost for five instances of the experiment. The standard error of this cost measurement is used as the uncertainty estimate for the cost. This uncertainty is not incorporated into the cost itself, but is fed directly to M-LOOP alongside the cost. M-LOOP is then able to estimate errors on the parameter landscapes [e.g., Fig.~\ref{fig:TOData}(b)] using these input uncertainties.

While vortices are the primary spurious effects we are interested in minimizing, density excitations (sound) may also result depending on the stirring profile~\cite{Kiehn_2022}. This information is not readily extracted from the TOF images, although it may affect the regularity of the fringe pattern. Minimizing sound excitation may be an additional refinement in future implementations of this approach.

Equation~(\ref{eqn:CostTargeting}) is minimized when the winding number is exactly $N_T$ and no vortices are present in the system, resulting in optimization to the target. Equation~(\ref{eqn:Costmaximization}) is minimized for the largest possible winding number given a minimized vortex number, thus maximizing the PC. The weights on each term were chosen such that any spurious vortices are weighted equally to the error in the winding for Eq.~(\ref{eqn:CostTargeting}) or to how large the winding becomes in Eq.~(\ref{eqn:Costmaximization}). This ensures that the best ``score" is assigned not only in achieving the target or maximizing the current, but also when spurious vortices are removed. When including the stirring time penalty, the factor of $10$ ensures that  that longer stirring times are heavily punished, to the same level of having ten spurious vortices for every stirring time $T$ near the upper limit $T_{\text{max}}$.

\section{\label{sec:TUOptimisation}Optimizations O.I and O.II: Variable stirring time}
We begin our investigation by fixing the stirring exponent $P=2$ (constant acceleration) and considering $T_{\text{max}}\to\infty$ to remove the stirring time penalty from Eqs.~(\ref{eqn:CostTargeting}) and~(\ref{eqn:Costmaximization}), meaning the stirring time is limited only by its upper bound. 
\subsection{Optimization O.I: Targeting}
Using Eq.~(\ref{eqn:CostTargeting}) as a cost function aims to realize a target winding number. This case is desirable for experiments investigating various initial conditions of a superfluid shear layer such as with the superfluid Kelvin-Helmholtz instability~\cite{Baggaley_2018, hernandez2023universality} or when fluxons~\cite{Oliinyk_2019,Oliinyk_2020} are experimentally initialized by stirring. A target winding number of $N_T=20$ is chosen. This choice was made since a winding of $20$ was already known to be achievable in the experiment though manually optimized stirring. The general process for optimization is described in Fig.~\ref{fig:ProcessFlow}. After 50 iterations, the learner was terminated, and the results of the procedure can be seen in Figure~\ref{fig:TOData}.

The cost vs run number is shown in Fig.~\ref{fig:TOData}(a) and illustrates how the learner evolved in choosing parameters and their effectiveness in experimentally acquiring a circulation of $N_T=20$. We note that in this scheme it is difficult to discern between large cost values due to the measured winding number being vastly different to the target or whether it is due to many spurious vortices. Generally, spurious vortices vary greatly shot to shot and so large cost values with larger uncertainties can be associated with large spurious vortex numbers. In contrast, the winding will be relatively stable shot to shot and data points associated with these defects will have lower uncertainty. The blue curve superimposed on the data tracks the running minimum cost. As can be seen, convergence is quickly achieved, justifying the cutoff run number of $50$ since we see no significant improvement in the cost or parameters. With these parameters, the cost landscape forms a four-dimensional region. Landscape curves, shown in Fig.~\ref{fig:TOData}(b), are computed as the cross sections of the cost landscape for each of the three parameters, chosen such that the other two parameters coincide with experimentally minimal costs. The optimum stirring parameters can be read directly from the global minima of the curves.

Figure~\ref{fig:TOData}(c) shows the results of stirring under the learner-proposed optimum parameters. For this particular image, no spurious vortices were observed while the winding count is $N_W = 19$. Averaging over five runs under these optimum parameters, it was found that the mean winding was $\overline{N}_W = 19.6\pm0.3$ with $\overline{N}_V = 0.2\pm0.2$ spurious vortices. The optimal stirring time $T$ is $760.6$-ms, near the middle of its bound range. The optimal removal time $T_R = 176.2$-ms is relatively short, near its lower bound. The coefficient parameter $\alpha$ exhibits multiple minima but has one clear global minimum at $0.786$-rad. Using these, it is possible to also deduce the final speed of the barrier from Eq.~\ref{eqn:StirProf1} to be $v_B=8.9\times10^{-3}$-rad/ms or equivalently $v_B=0.38$-mm/s at the inner radius of $32.25$-$\mu$m. The optimum parameters, along with the parameters for the subsequent optimizations, are summarized in Table~\ref{table:ParameterSummary}.
\subsection{\label{sec:CM} Optimization O.II: Maximization}
\begin{figure}[t!]
    \centering
\includegraphics[trim={0cm 5.2cm 0cm
0cm},clip,width=\hsize]{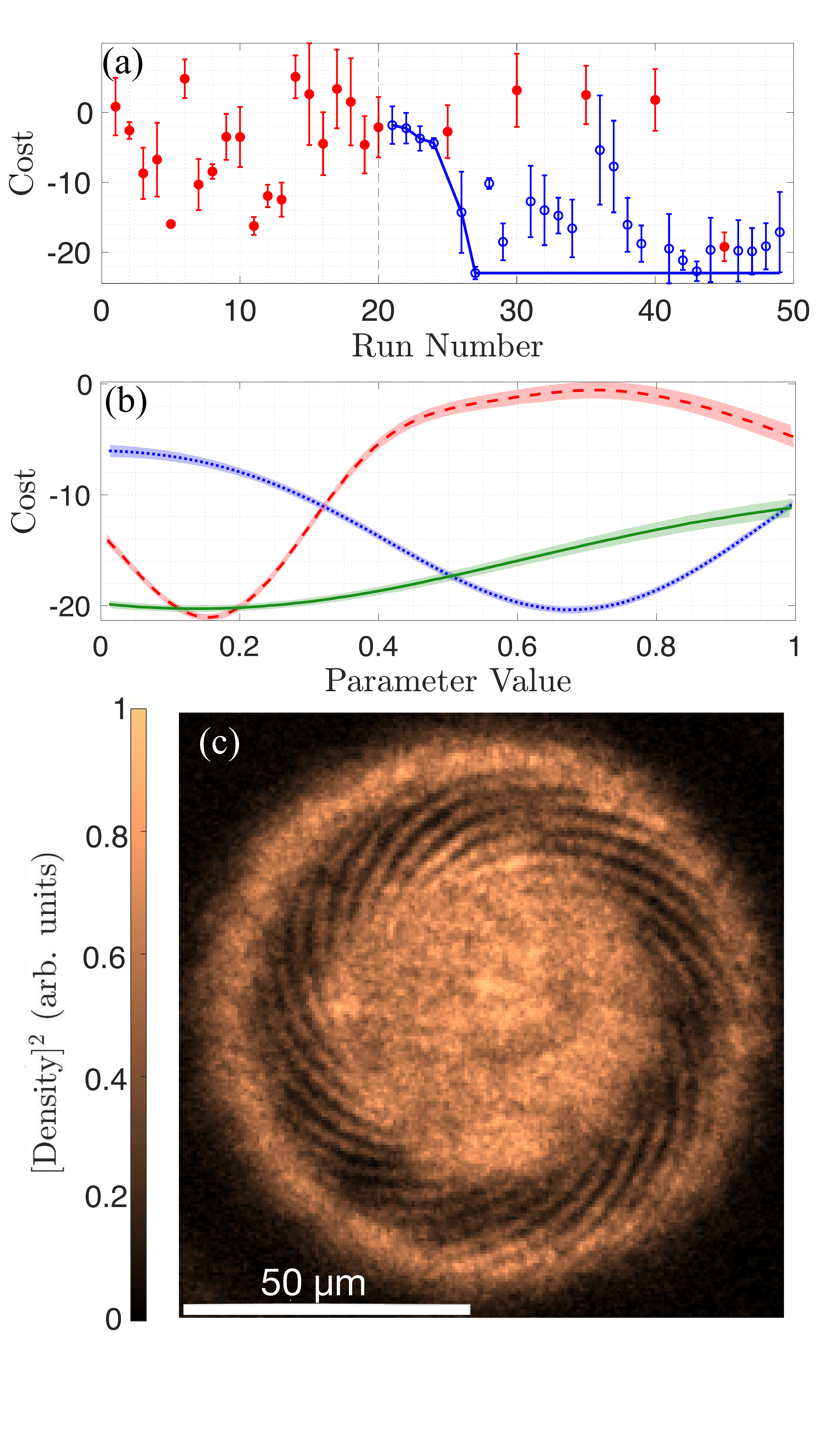}
\caption{M-LOOP output graphics for maximization of the winding number (O.II -- cost function given by Eq.~(\ref{eqn:Costmaximization})). (a) The cost vs the run number over 50 runs. (b) The predicted landscape showing the optimum parameter choices. The dashed (red) line represents the stirring time parameter $T$, the solid (green) line represents the removal time $T_R$, and the dotted (blue) line represents the angular coefficient $\alpha$. Uncertainties are indicated by the shaded regions. (c) $5$-ms TOF image under optimized parameters. This particular image shows 25 fringes.}
\label{fig:MaxData}
\end{figure}

Optimization O.II seeks to maximize the winding number while suppressing spurious vortices. This can be achieved by running M-LOOP with the cost function in Eq.~(\ref{eqn:Costmaximization}). Figure~\ref{fig:MaxData} shows the outcome of this optimization. There is immediately a clear and steady decrease in cost after the final training run, leading to a plateau in the cost for the final runs indicated by the minimum cost curve in Fig.~\ref{fig:MaxData}(a). This behavior suggests convergence to an optimal parameter set.

The parameter landscapes show the optimized parameter choices as clear global minima of the curves. Unlike the targeting optimization, all curves contain a single minima. The optimal stirring protocol applies aggressive movement of the barrier, indicated by the large $\alpha=2.36$-rad, but for a relatively short stirring time at $T=335.3$-ms and removal time $T_R=15.2$-ms. These parameters correspond to a final barrier speed of $v_B=1.0\times10^{-2}$-rad/ms or $v_B=0.44$-mm/s at a radius of $32.25$-$\mu$m. An example TOF image using the optimized parameters is shown in Fig.~\ref{fig:MaxData}(c), which contains $25$ windings. The parameters result in $\overline{N}_W = 25.8\pm0.4$, with $\overline{N}_V = 0.2\pm0.2$ over five experimental runs.
\subsection{\label{sec:Discussion1}Remarks on the optimizations}
 Considering first Fig.~\ref{fig:TOData}(a) and Fig.~\ref{fig:MaxData}(a), we observe that during training runs the computed cost and uncertainty vary significantly from run to run. The concept behind this random training is to walk through as large a portion of the parameter space as possible before the optimization effectively begins. Doing so ensures that the Gaussian process regression is more accurately able to produce meaningful landscape curves~\cite{Wigley_2016}. Following the training, we observe a rapid stagnation of the cost in both optimization cases. This is indicated by the minimum cost curves superimposed on Fig.~\ref{fig:TOData}(a) and Fig.~\ref{fig:MaxData}(a), which show a drop after training to a sudden plateau suggesting convergence. 

For maximization O.II, the optimal parameters can be interpreted as those which reach the upper limit for the winding (flow speed) of the persistent current before the shedding of vortices from the stirring barrier cannot be avoided, i.e., the critical velocity has been reached~\cite{Reeves_2015}. Applying  Eq.~(\ref{eqn:MeanVelCalc}) [Appendix~\ref{sec:VelocityEstimates}] using the final winding numbers for both the targeting and maximization case gives a flow speed of $v_{\text{T}}(R)\approx347$-$\mu$m/s and $v_{\text{M}}(R)\approx457$-$\mu$m/s respectively (at the inner ring radius $R=32.25~\mu$m). Equivalently, this gives the critical velocity of the experimental system to be $v_{\text{M}}\approx0.35 c_s$ where $c_s\approx1.3$-mm/s is the two-dimensional (2D) speed of sound for the system [see Appendix B, Eq.~(\ref{eqn:speedOfSound})]. This critical velocity measurement is consistent with previous estimates of vortex shedding due to flow past an obstacle~\cite{raman1999evidence,Stiesberger_2000,Kwon_2015}, suggesting the current is limited by features or roughness in the optical potential and that the machine learner has optimized the stirring process to this limit. We also note that the final barrier and fluid speeds for O.I do not coincide, as shown in Table~\ref{table:ParameterSummary}, indicating that the fluid is out of equilibrium with the barrier for this stirring protocol. This behavior is discussed in more detail in Sec.~\ref{sec:Conclusion}.

Overall, the machine learner is able to optimize two different stirring protocols for different final PC conditions. Given the constant angular acceleration used in the stirring profile here, an experimenter might opt to maximize the current via stirring using the same profile as O.I for longer times, given that this is the most straightforward approach. Remarkably, the machine learner finds a different and more efficient approach to stirring: aggressively stirring for a shorter period of time. 
\begin{figure}[t!]
    \centering
\includegraphics[trim={0cm 5.2cm 0cm
0cm},clip,width=\hsize]{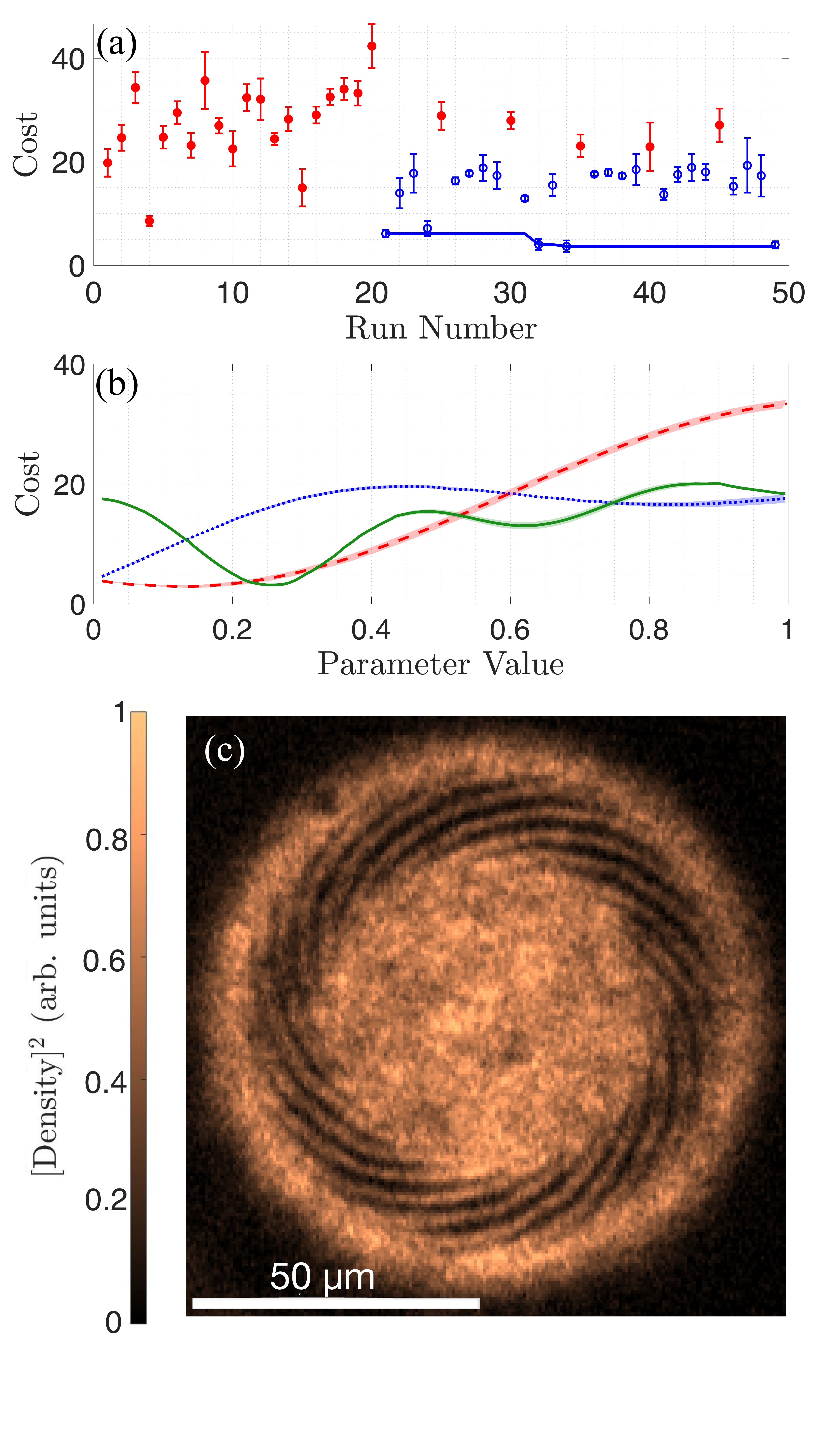}
\caption{M-LOOP output graphics for time-limited targeting a winding number of $N_T=20$ windings (O.III -- cost function given by Eq.~(\ref{eqn:CostTargeting})). (a) The cost vs the run number over 50 runs. The color convention is consistent with Fig.~\ref{fig:TOData}. (b) The predicted landscape showing the optimum parameter choices. The dashed (red) line represents the stirring time $T$ landscape while the dotted (blue) line represents the acceleration coefficient $\alpha$. Here, the solid (green) line represents the exponent parameter $P$. Uncertainties are indicated by the shaded regions. (c) $5$ ms TOF image under optimized parameters. This particular image shows 18 fringes.}
\label{fig:TLTargetData}
\end{figure}
\section{\label{sec:MinimumTime}Optimizations O.III and O.IV: Minimum stirring time}
Motivated by the above results, we next repeat the previous optimizations while attempting to minimize the stirring time. The learner now has to satisfy the desired winding conditions with a cost penalty for long stirring times $T$. We no longer hold $P=2$, in Eqs.~(\ref{eqn:Costmaximization}) and (\ref{eqn:CostTargeting}), instead allowing it to be a parameter optimized by the learner. This extra flexibility is allowed in order to compensate for the added complexity in the cost function, giving the learner more control over the barrier motion. $T_{\text{max}}=1500$-ms is now the maximum allowed stirring time, which is consistent with the upper bound on the stirring time $T$. Since the cost was observed to vary weakly with the barrier removal time (see Figs.~\ref{fig:TOData} and \ref{fig:MaxData}) the removal time is fixed to $T_R=100$-ms, simplifying the cost functions and aiding convergence.

\subsection{\label{sec:TRTO}Optimization O.III: Targeting}

Figure~\ref{fig:TLTargetData} shows the outcomes from target optimization subject to a time cost imposed by Eq.~(\ref{eqn:CostTargeting}). The stirring time curve in Fig.~\ref{fig:TLTargetData}(b) shows a global minimum near $240$-ms, which is close to its lower boundary of $100$-ms, indicating that only short stirring times are required to achieve the desired winding target. The power parameter also has its global minimum at its lower bound value $P=1$ and the associated coefficient $\alpha=2.64$-rad. This results in an instantaneous and constant angular velocity of $6.6\times10^{-3}$-rad/ms, corresponding to $0.28$-mm/s at the inner radius of $32.25$-$\mu$m. Spurious vortices were more prevalent in this optimization, as shown in the example image, Fig.~\ref{fig:TLTargetData}(c). Here we also observe 18 windings. Averaging over five separate images, a winding number of $N_W=18.4\pm0.4$ is observed with $N_V=1.0\pm0.3$ spurious vortices.
\subsection{\label{sec:TRCM}Optimization O.IV: Time-limited maximization}
\begin{figure}[t!]
    \centering
\includegraphics[trim={0cm 5.2cm 0cm
0cm},clip,width=\hsize]{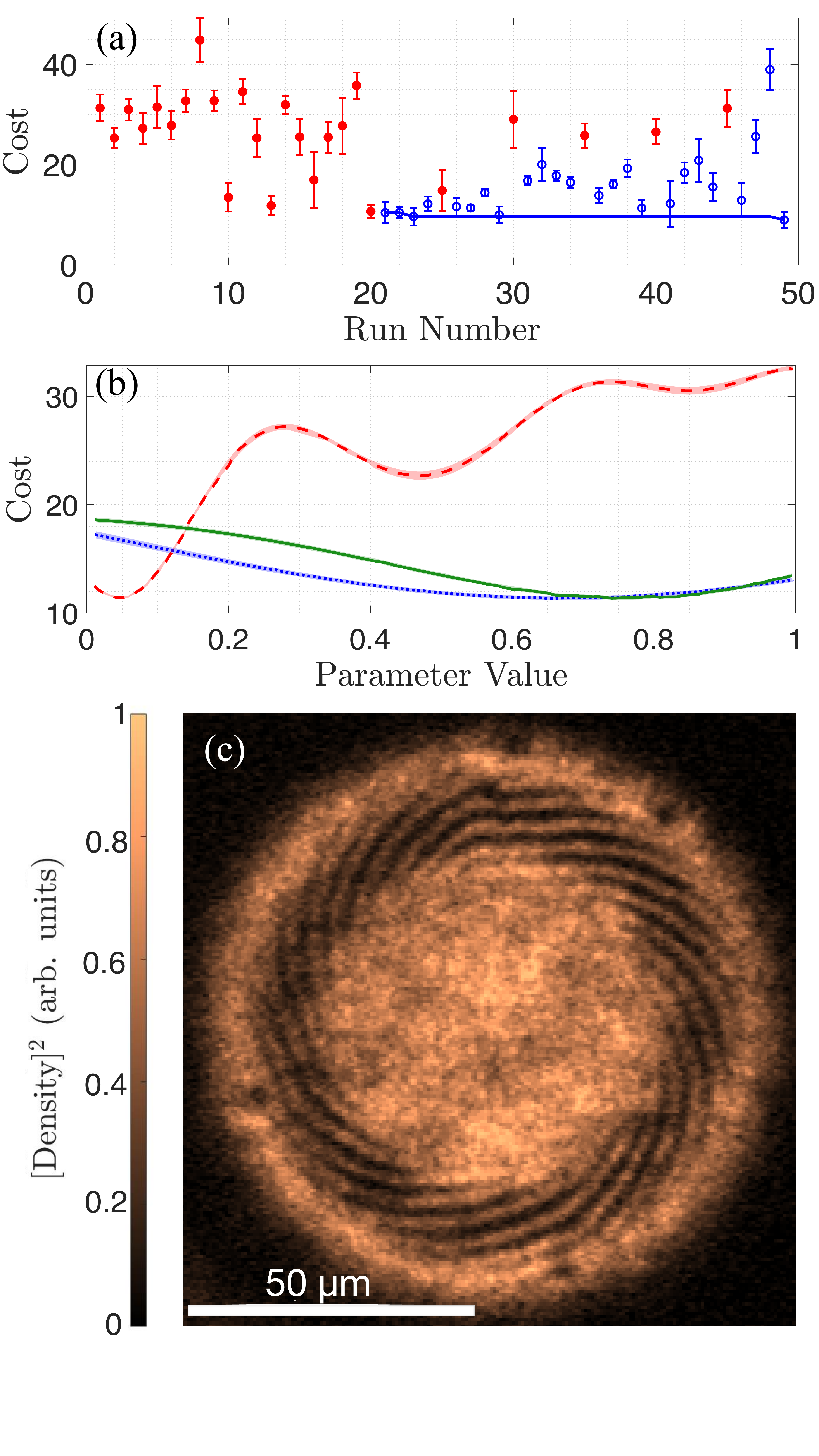}
\caption{M-LOOP output graphics for time-limited maximization of the windings number (Optimization IV -- cost function given by Eq.~(\ref{eqn:Costmaximization})). (a) The cost vs the run number over 50 runs. (b) The predicted landscape shows the optimum parameter choices. The dashed (red) line represents the stirring time $T$ landscape while the dotted (blue) line represents the acceleration coefficient $\alpha$. Here the solid (green) line represents the exponent parameter $P$. Uncertainties are indicated by the shaded regions. (c) $5$ ms TOF image under optimized parameters. This particular image shows 18 fringes.}
\label{fig:TLMaxData}
\end{figure}

Figure~\ref{fig:TLMaxData} shows the outcome of trying to maximize the winding by employing Eq.~(\ref{eqn:Costmaximization}) with a time cost for M-LOOP. The learner was able to find optimized parameters as indicated by the clear minima in the landscape curves of Fig.~\ref{fig:TLMaxData}(b). All global minima lie away from the boundaries of the parameters in this optimization. The stirring is highly non-linear, with a value of $P=3.5$. The stirring time was optimized to $T=163.5$-ms suggesting that the learner correctly identified the cost function and time restriction. Accompanying the short, non-linear stirring is a large, $\alpha=8.22$-rad, coefficient contributing to the highly aggressive stirring. In this case, the final barrier speed is $v_B=2.5\times10^{-2}$-rad/ms or $1.1$-mm/s at a radius of $32.25$-$\mu$m. Under this proposed optimal stirring, the resulting BEC looks similar to Fig.~\ref{fig:TLMaxData}(c) which contains five spurious vortices but only 18 windings. An average winding number of $\overline{N}_W = 18.4\pm 0.4$ with $\overline{N}_V = 4.5\pm 0.5$ spurious vortices is observed overall. Likely this smaller winding, relative to O.II, is due to the trade-off placed in the cost function weights between the winding and the stirring time, which are necessary to enable the fast creation of the PC. 
\subsection{\label{sec:Discussion2}Remarks on the time-limited optimizations}

The time series of the cost vs run number for O.III in Fig.~\ref{fig:TLTargetData}(a) shows a small dip followed by stagnation within ten machine learner controlled runs. Both cases exhibit the characteristic random cost distribution during training, and the effectiveness of said training is highlighted by the smoothness of the parameter landscape curves in Figs.~\ref{fig:TLTargetData}(b) and \ref{fig:TLMaxData}(b). In both cases, the parameter landscape curves have minima near the lower boundary of the stirring time $T$. This suggests that the learner correctly includes the added restriction on the stirring time. 

Observing the time series for both O.III and O.IV, Figs.~\ref{fig:TLTargetData}(a) and \ref{fig:TLMaxData}(a) respectively, we see the cost vs run number in both cases stagnates rapidly, suggesting fast convergence. However, the convergence in winding number is exactly the same between the two cases. Since both cases share $N_W=18.4\pm 0.4$, both have a flow speed of $v(R)\approx$-$\mu$m/s or $v\approx.25 c_s$ (at $R=32.25$-$\mu$m). For O.III, this is close to the target. For O.IV however, this flow rate is far from the critical velocity measured in O.II. The inability of the learner to optimize the stirring as desired arises from the fact that the cost function is designed with a new weighted punishment term for the stirring time, introducing a trade-off between maximum winding and shortest stirring.

\section{\label{sec:Conclusion}Conclusion and Outlook}
The first aim of this paper was to realize optimal stirring parameters for creating PCs with a stirring barrier. The machine learning algorithm was successful in this endeavor, creating PCs for each of the optimizations explored, while minimizing spurious vortices~\footnote{We note that the lower winding numbers in O.III and O.IV result from the strong weighting on the stirring time punishment term in the cost functions. This observation highlights the trade-offs introduced in the cost functions which are likely the limiting factors preventing more complete optimization in more complicated scenarios.}. While our initial hypothesis was that there would be a unique approach to optimal stirring, the results depended greatly on the choice of cost function and the constraints chosen. In the case of variable stirring time without penalty (O.I and O.II), the learner chose dramatically different stirring times, stirring gently for a long time (O.I) and stirring aggressively for a short time (O.II). For the variable time optimizations (O.III and O.IV), the choice of parameters is even more distinct. O.III found the best protocol was to instantaneously accelerate the barrier to move with constant angular velocity, while O.IV found that aggressive stirring with a highly non-linear profile, $P=3.5$, was optimal. 

\begin{table}[t]
\resizebox{8.6cm}{!}{%
\begin{tabular}{|c||c|c|c|c|c|c|}
\hline
Optimization & $T$~(ms) & $T_R$~(ms) &  $\alpha$ (rad) & $P$ & $v$/$c_s$ & $v_B$/$c_s$\\ \hline
\hline
O.I & 721 & 176 & 0.786 & 2 & 0.27 & 0.38\\ \hline
O.II & 335 & 15.2 & 2.36 & 2 & 0.35 & 0.34\\ \hline
O.III & 240 & 100 & 2.64 & 1 & 0.25 & 0.22\\ \hline
O.IV & 164 & 100 & 8.22  & 3.5 & 0.25 & 0.84\\ \hline
\end{tabular}%
}
\caption{Summary of optimized parameter values for the  optimization cases O.I (targeting), O.II (maximization), O.III (time-limited targeting), and O.IV (time-limited maximization). $v$ is the PC flow speed in the middle of the ring, while $c_s\approx1300$-$\mu$m/s is the speed of sound. The final speed of the barrier is indicated by $v_B$.}
\label{table:ParameterSummary}
\end{table}

These results suggest that while the learner can successfully control stirring for desired results and minimum cost, the problem of determining an optimal stirring profile is under-constrained by the chosen parameters. We take this result to imply that stirring to create a PC is a robust process, and can be successfully achieved with a variety of stirring profiles. From a practical perspective, the relatively counter-intuitive approach of aggressive stirring for a short time at constant speed (O.III) may be the best approach, since this minimizes atom loss due to the finite lifetime of the BEC.
Another interesting feature of these optimizations is that there are some out-of-equilibrium dynamics between the fluid and barrier. This is seen through the difference in final barrier speeds relative to the superfluid velocity. For example, O.IV resulted in a final barrier speed of $v_B=0.84c_s$ for a final superfluid speed of $v=0.25c_s$, but relied on a short stirring time of $164$-ms. These results suggest counter-adiabatic stirring protocols may be advantageous, provided the barrier beam is removed at the right time. Previous work has explored the possibility of spatial rapid adiabatic passage for engineering angular momentum states in a short time~\cite{Menchon_2016,Guery_2019}. Future extension of similar approaches to stirring of persistent currents may be a fruitful area of both theoretical and experimental future work.

Overall, the successful optimizations demonstrate the suitability of using machine learners for the creation and control of PCs through stirring, albeit with no universal stirring protocol. These results suggest useful future applications of the algorithm. Since PCs establish a basis for atom interferometry~\cite{Pandey_2019,Edwards_2013,Gross2010}, having high levels of control over PCs using machine learning can be desirable for engineering atom interferometers~\cite{Levy2007,Packard_1992}. Beyond interferometry, it is possible to apply this algorithm to prepare PC states with specified winding numbers for studies in quantum turbulence, such as determining the dynamics of the Kelvin-Helmholtz instability~\cite{Baggaley_2018,hernandez2023universality}. Atom inertial sensors have used PCs as a source of synthetic rotation for calibration~\cite{woffinden2022viability}, and this method presents a precise way of creating these references. The general structure of the feedback loop described could also be adapted to optimize alternative methods for generating PCs, such as phase imprinting~\cite{del2022imprinting}.

\section{\label{sec:Acknowledgments}Acknowledgements}
We thank M.~Reeves and M.~Christenhusz for useful discussions and help with computational work. This research was supported by the Australian Research Council (ARC) Centre of Excellence for Engineered Quantum Systems (EQUS, CE170100009). S.~S~acknowledges the support of an Australian Government Research and Training Program Scholarship. G.~G.~acknowledges the support of ARC Discovery Projects Grant No. DP200102239. T.~W.~N.~acknowledges the support of Australian Research Council Future Fellowship No. FT190100306.

\appendix
\section{\label{sec:Imaging}Image Processing}
 Images of the BEC are produced using Faraday imaging~\cite{GajdaczMiroslav2013NFio,Wilson_2015}. Automatic image processing is used to extract the winding number and vortex number from the Faraday images. Vortices become directly resolvable in TOF as clusters of dark pixels, corresponding to density dips, which means that detection of such sites can be treated analogously to blob detection in general image processing~\cite{bauckhage2015k,Rakonjac2016,Guillaume2020}. The use of this algorithm ensures successful detection of vortices outside of the fringe region but tends to count dark fringes as vortex objects as well. To avoid this issue, the algorithm is restricted to operate only outside the fringe region through masking while the fringe region is analyzed using a Fourier signal method discussed in more detail below. Specifically, we restrict the vortex detection to the outer ring where stirring occurs. 

\begin{figure}[t]
    \centering
    \includegraphics[trim={13.3cm 2cm 18cm
2cm},clip,width=\hsize]{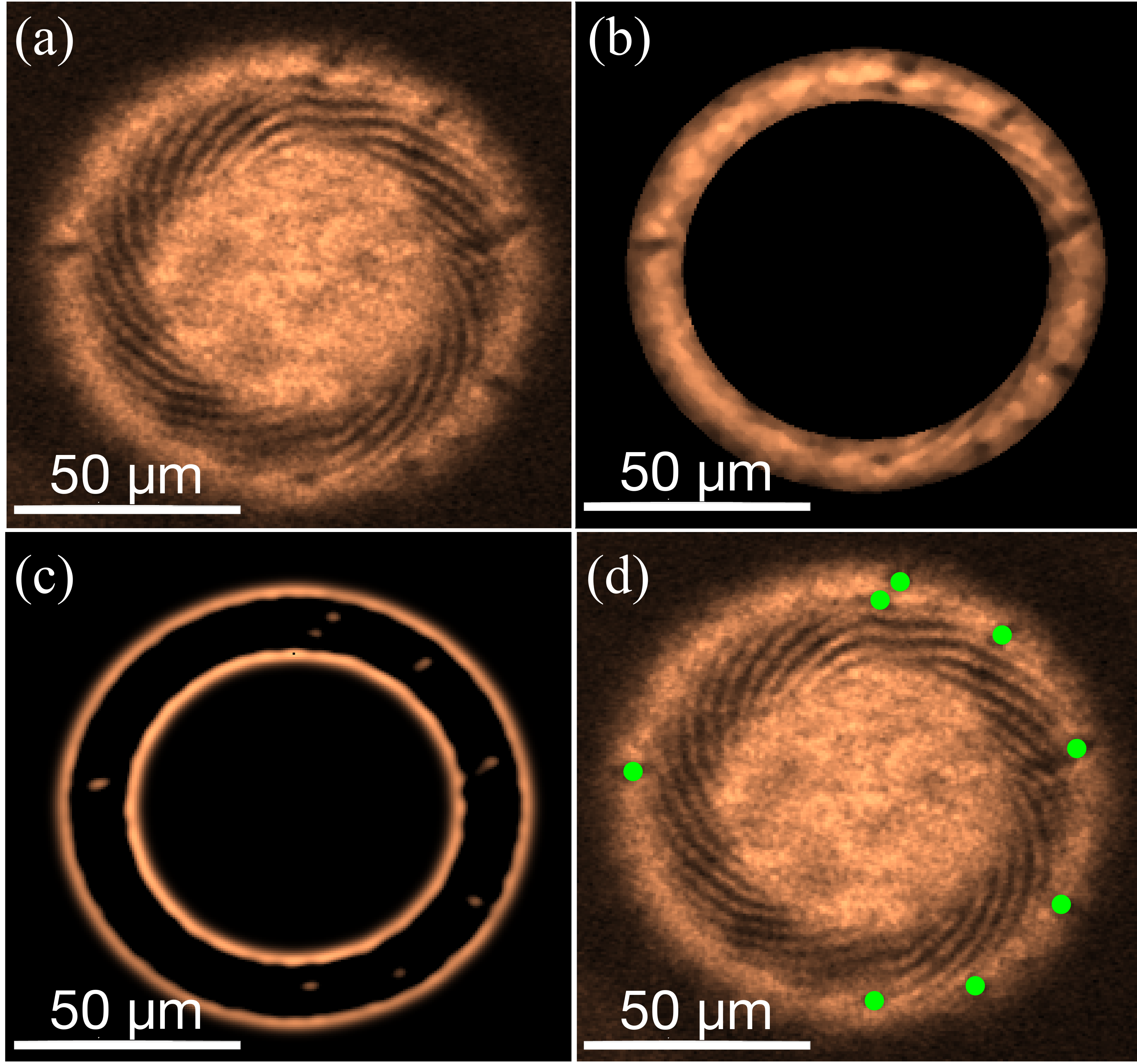}
    \caption{Processing of a TOF image using the Gaussian blob algorithm. (a) The original image. (b) Masked image to remove the fringe region. The processed image is also smoothed using a Gaussian filter. (c) Laplacian of the masked and filtered image. Thresholding of the image is now possible and the sharp ring edges can be masked out. (d) Detected vortex positions (green solid circles) superimposed on the original image. Additional details can be found in Refs.~\cite{bauckhage2015k,Rakonjac2016,Guillaume2020}. Note that due to the masking of the fringe region, vortices in this region are not detected despite being clearly present as evident in the warping and forking of the fringes.}
    \label{fig:VortDetEx}
\end{figure}

Vortices may still be present in the fringe region and can be observed as either a warping or a fork in a fringe which separates into two fringes, most clearly observable in Fig.~\ref{fig:VortDetEx}(a). This results in an increased standard deviation when estimating the winding number as described below. The fringe number standard deviation is simply added to the vortex number $N_V$ when computing the average cost for the set of five runs. Although this potentially overestimates the number of vortices in the fringe region by mixing the standard deviation in achieved winding number into $N_V$, we find that this approach successfully minimizes spurious vortices in both the masked and interference fringe regions. Note that any spurious vortices in the interior of the reference condensate are ignored by the algorithm since they are not caused by the stirring process.

\begin{figure}[t!]
    \centering
    \includegraphics[trim={0cm 2cm 0cm
3cm},clip,width=\hsize]{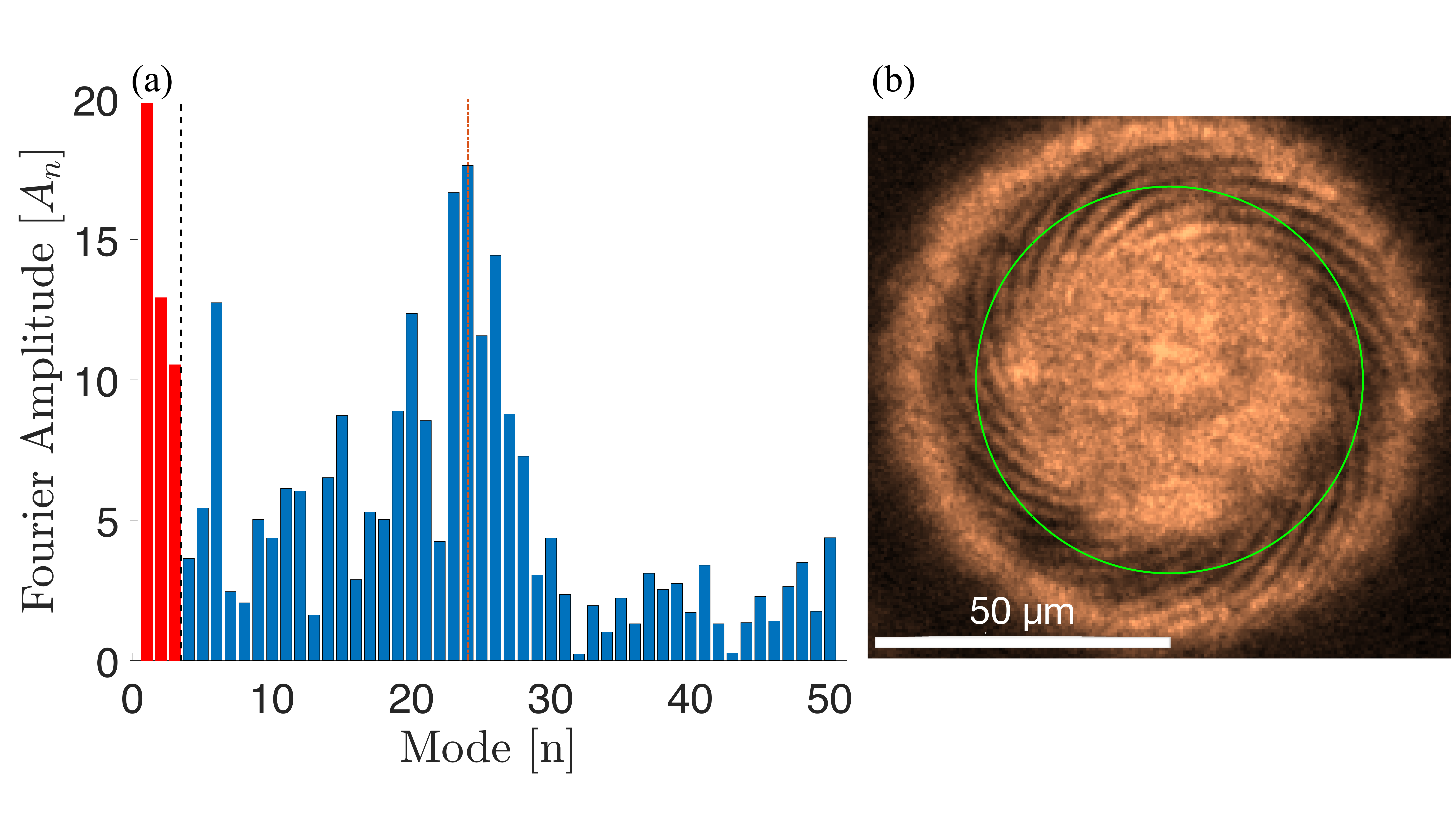}
     \caption{Example of the processing of a TOF image using the winding counting algorithm. (a) Fourier amplitude for mode numbers 1--50. The largest amplitude mode is taken to be the winding number. Mode numbers at or under the cutoff mode of $n=3$ are indicated by the black dashed vertical line and are highlighted in red. This cutoff is introduced since these modes have a large amplitude but do not reflect the fringe number. The investigation does not require modes close to $n=3$ to be detected. The largest amplitude has a mode number of $24$, indicated by the orange, dot-dashed line. (b) Color coded image highlighting the region of interest (green ring) considered in the example, superimposed on the original image. Note that only one small region of interest is shown here, while in practice 1 similar rings of varying radii are considered. These 1 rings overlap slightly since their inner radii are separated by $0.25$-$\mu$m while their thickness remains fixed at $1$-$\mu$m. Over the 1 rings for the example image the mean fringe number is $24.8\pm0.4$ where the uncertainty is estimated as the standard error of the 1 ring measurements. The observed count in this particular image is $24$ fringes.}
    \label{fig:DFTExample}
\end{figure}

A Fourier signal algorithm can be used for the counting winding number and its standard deviation for the aforementioned estimate of $N_V$. We begin by considering the fringe region as indicated by the solid green line in Fig.~\ref{fig:DFTExample}(b). The azimuthal Fourier transform is determined and the mode with the largest Fourier amplitude is identified as corresponding to the winding number for the selected region. A cutoff mode of $n=3$ is introduced since the signal of these modes is generally large, leading to erroneous detection of winding numbers that do not correspond to the observed spiral pattern. We found this cutoff value sufficient for reducing this effect. 
 

The spiraling nature of the fringes requires some care in considering how to best extract the winding number from the discrete Fourier transform. The solution is to subdivide the region into 1 overlapping rings of 1-$\mu$m thickness. These smaller rings are set apart via their inner and outer radii being translated by 0.25-$\mu$m. This choice allows the entire fringe region to be spanned by a series of thin rings over which the accuracy of the detection is improved. The standard deviation of the winding number across the rings is used to estimate the number of vortices in the fringe region. Finally, in order to extract a useful measure of winding number, the mean of the mode number with the largest amplitude is selected which need not be an integer. Figure~\ref{fig:DFTExample}(a) shows the Fourier amplitudes for the considered modes for the small ring region considered in the image in Fig.~\ref{fig:DFTExample}(b). 

\section{\label{sec:VelocityEstimates}Estimating Flow Speed from Winding Numbers}
It is possible to estimate the speed of the PCs generated by these different stirring processes via the velocity-phase relationship in a supercurrent:
\begin{equation}
    \vec{v}(\vec{r},t)=\frac{\hbar}{m}\nabla \phi(\vec{r},t). \label{eqn:PhaseVelocityRel}
\end{equation}
Assuming that the winding is in the azimuthal direction,  $\hat{\theta}$, due to the ring geometry, any radial derivatives of the phase are zero. The PC should not decay making the speed independent of time. Therefore, the flow speed can be computed via:
\begin{equation}
    \vec{v}(r)=\frac{\hbar}{m}\frac{1}{r}N_W\hat{\theta}.\label{eqn:MeanVelCalc}
\end{equation}
The speed of sound in the quasi-2D BEC system can be estimated for comparison using the relation:
\begin{equation}
    c_s=\sqrt{\frac{\mu_{\text{2D}}}{m}}, \label{eqn:speedOfSound}
\end{equation}
where-$\mu_{\text{2D}}$ is the effective 2D chemical potential of the persistent current ring region.-$\mu_{\text{2D}}$ can be deduced from the condensed atom number in the ring $N_c\approx 9.6\times10^5$ along with the inner and outer radii of the ring, $r_\text{I}=36$-$\mu$m and $r_\text{O}=50$-$\mu$m respectively, using the following relation~\cite{Gauthier_2019}:
\begin{equation}
    \mu_{\text{2D}}=\frac{2}{5}\left(\frac{3}{2}\frac{gN_c\sqrt{m\omega_z^2}}{\pi(r_{\text{O}}^2-r_\text{I}^2)}\right)^{2/3}. \label{eqn:chemPot}
\end{equation}
Using a vertical trapping frequency of $\omega_z=2\pi\times 108$ Hz and the $s$-wave scattering length $a_s=98.13a_0$ to compute the effective 2D interaction parameter $g=4\pi\hbar^2a_s/m$, we compute that-$\mu_{\text{2D}}\approx2.5\times10^{-31}$ J/kg and therefore $c_s\approx1.3$-mm/s. 
\bibliography{references}

\begin{thebibliography}{62}%
\makeatletter
\providecommand \@ifxundefined [1]{%
 \@ifx{#1\undefined}
}%
\providecommand \@ifnum [1]{%
 \ifnum #1\expandafter \@firstoftwo
 \else \expandafter \@secondoftwo
 \fi
}%
\providecommand \@ifx [1]{%
 \ifx #1\expandafter \@firstoftwo
 \else \expandafter \@secondoftwo
 \fi
}%
\providecommand \natexlab [1]{#1}%
\providecommand \enquote  [1]{``#1''}%
\providecommand \bibnamefont  [1]{#1}%
\providecommand \bibfnamefont [1]{#1}%
\providecommand \citenamefont [1]{#1}%
\providecommand \href@noop [0]{\@secondoftwo}%
\providecommand \href [0]{\begingroup \@sanitize@url \@href}%
\providecommand \@href[1]{\@@startlink{#1}\@@href}%
\providecommand \@@href[1]{\endgroup#1\@@endlink}%
\providecommand \@sanitize@url [0]{\catcode `\\12\catcode `\$12\catcode
  `\&12\catcode `\#12\catcode `\^12\catcode `\_12\catcode `\%12\relax}%
\providecommand \@@startlink[1]{}%
\providecommand \@@endlink[0]{}%
\providecommand \url  [0]{\begingroup\@sanitize@url \@url }%
\providecommand \@url [1]{\endgroup\@href {#1}{\urlprefix }}%
\providecommand \urlprefix  [0]{URL }%
\providecommand \Eprint [0]{\href }%
\providecommand \doibase [0]{https://doi.org/}%
\providecommand \selectlanguage [0]{\@gobble}%
\providecommand \bibinfo  [0]{\@secondoftwo}%
\providecommand \bibfield  [0]{\@secondoftwo}%
\providecommand \translation [1]{[#1]}%
\providecommand \BibitemOpen [0]{}%
\providecommand \bibitemStop [0]{}%
\providecommand \bibitemNoStop [0]{.\EOS\space}%
\providecommand \EOS [0]{\spacefactor3000\relax}%
\providecommand \BibitemShut  [1]{\csname bibitem#1\endcsname}%
\let\auto@bib@innerbib\@empty
\bibitem [{\citenamefont {Leggett}(2001)}]{leggett2001bose}%
  \BibitemOpen
  \bibfield  {author} {\bibinfo {author} {\bibfnamefont {A.~J.}\ \bibnamefont
  {Leggett}},\ }\bibfield  {title} {\bibinfo {title} {Bose-{E}instein
  condensation in the alkali gases: Some fundamental concepts},\ }\href@noop {}
  {\bibfield  {journal} {\bibinfo  {journal} {Reviews of Modern Physics}\
  }\textbf {\bibinfo {volume} {73}},\ \bibinfo {pages} {307} (\bibinfo {year}
  {2001})}\BibitemShut {NoStop}%
\bibitem [{\citenamefont {Ryu}\ \emph {et~al.}(2007)\citenamefont {Ryu},
  \citenamefont {Andersen}, \citenamefont {Clade}, \citenamefont {Natarajan},
  \citenamefont {Helmerson},\ and\ \citenamefont
  {Phillips}}]{ryu2007observation}%
  \BibitemOpen
  \bibfield  {author} {\bibinfo {author} {\bibfnamefont {C.}~\bibnamefont
  {Ryu}}, \bibinfo {author} {\bibfnamefont {M.~F.}\ \bibnamefont {Andersen}},
  \bibinfo {author} {\bibfnamefont {P.}~\bibnamefont {Clade}}, \bibinfo
  {author} {\bibfnamefont {V.}~\bibnamefont {Natarajan}}, \bibinfo {author}
  {\bibfnamefont {K.}~\bibnamefont {Helmerson}},\ and\ \bibinfo {author}
  {\bibfnamefont {W.~D.}\ \bibnamefont {Phillips}},\ }\bibfield  {title}
  {\bibinfo {title} {Observation of persistent flow of a {B}ose-{E}instein
  condensate in a toroidal trap},\ }\href@noop {} {\bibfield  {journal}
  {\bibinfo  {journal} {Phys. Rev. Lett.}\ }\textbf {\bibinfo {volume} {99}},\
  \bibinfo {pages} {260401} (\bibinfo {year} {2007})}\BibitemShut {NoStop}%
\bibitem [{\citenamefont {Neely}\ \emph {et~al.}(2013)\citenamefont {Neely},
  \citenamefont {Bradley}, \citenamefont {Samson}, \citenamefont {Rooney},
  \citenamefont {Wright}, \citenamefont {Law}, \citenamefont
  {Carretero-Gonz{\'a}lez}, \citenamefont {Kevrekidis}, \citenamefont {Davis},\
  and\ \citenamefont {Anderson}}]{neely2013characteristics}%
  \BibitemOpen
  \bibfield  {author} {\bibinfo {author} {\bibfnamefont {T.~W.}\ \bibnamefont
  {Neely}}, \bibinfo {author} {\bibfnamefont {A.~S.}\ \bibnamefont {Bradley}},
  \bibinfo {author} {\bibfnamefont {E.~C.}\ \bibnamefont {Samson}}, \bibinfo
  {author} {\bibfnamefont {S.~J.}\ \bibnamefont {Rooney}}, \bibinfo {author}
  {\bibfnamefont {E.~M.}\ \bibnamefont {Wright}}, \bibinfo {author}
  {\bibfnamefont {K.~J.~H.}\ \bibnamefont {Law}}, \bibinfo {author}
  {\bibfnamefont {R.}~\bibnamefont {Carretero-Gonz{\'a}lez}}, \bibinfo {author}
  {\bibfnamefont {P.~G.}\ \bibnamefont {Kevrekidis}}, \bibinfo {author}
  {\bibfnamefont {M.~J.}\ \bibnamefont {Davis}},\ and\ \bibinfo {author}
  {\bibfnamefont {B.~P.}\ \bibnamefont {Anderson}},\ }\bibfield  {title}
  {\bibinfo {title} {Characteristics of two-dimensional quantum turbulence in a
  compressible superfluid},\ }\href@noop {} {\bibfield  {journal} {\bibinfo
  {journal} {Phys. Rev. Lett.}\ }\textbf {\bibinfo {volume} {111}},\ \bibinfo
  {pages} {235301} (\bibinfo {year} {2013})}\BibitemShut {NoStop}%
\bibitem [{\citenamefont {Amico}\ \emph {et~al.}(2021)\citenamefont {Amico},
  \citenamefont {Boshier}, \citenamefont {Birkl}, \citenamefont {Minguzzi},
  \citenamefont {Miniatura}, \citenamefont {Kwek}, \citenamefont {Aghamalyan},
  \citenamefont {Ahufinger}, \citenamefont {Anderson}, \citenamefont {Andrei}
  \emph {et~al.}}]{amico2021roadmap}%
  \BibitemOpen
  \bibfield  {author} {\bibinfo {author} {\bibfnamefont {L.}~\bibnamefont
  {Amico}}, \bibinfo {author} {\bibfnamefont {M.}~\bibnamefont {Boshier}},
  \bibinfo {author} {\bibfnamefont {G.}~\bibnamefont {Birkl}}, \bibinfo
  {author} {\bibfnamefont {A.}~\bibnamefont {Minguzzi}}, \bibinfo {author}
  {\bibfnamefont {C.}~\bibnamefont {Miniatura}}, \bibinfo {author}
  {\bibfnamefont {L.-C.}\ \bibnamefont {Kwek}}, \bibinfo {author}
  {\bibfnamefont {D.}~\bibnamefont {Aghamalyan}}, \bibinfo {author}
  {\bibfnamefont {V.}~\bibnamefont {Ahufinger}}, \bibinfo {author}
  {\bibfnamefont {D.}~\bibnamefont {Anderson}}, \bibinfo {author}
  {\bibfnamefont {N.}~\bibnamefont {Andrei}}, \emph {et~al.},\ }\bibfield
  {title} {\bibinfo {title} {Roadmap on {A}tomtronics: {S}tate of the art and
  perspective},\ }\href@noop {} {\bibfield  {journal} {\bibinfo  {journal} {AVS
  Quantum Science}\ }\textbf {\bibinfo {volume} {3}},\ \bibinfo {pages}
  {039201} (\bibinfo {year} {2021})}\BibitemShut {NoStop}%
\bibitem [{\citenamefont {Amico}\ \emph {et~al.}(2022)\citenamefont {Amico},
  \citenamefont {Anderson}, \citenamefont {Boshier}, \citenamefont {Brantut},
  \citenamefont {Kwek}, \citenamefont {Minguzzi},\ and\ \citenamefont {von
  Klitzing}}]{Amico_2022}%
  \BibitemOpen
  \bibfield  {author} {\bibinfo {author} {\bibfnamefont {L.}~\bibnamefont
  {Amico}}, \bibinfo {author} {\bibfnamefont {D.}~\bibnamefont {Anderson}},
  \bibinfo {author} {\bibfnamefont {M.}~\bibnamefont {Boshier}}, \bibinfo
  {author} {\bibfnamefont {J.-P.}\ \bibnamefont {Brantut}}, \bibinfo {author}
  {\bibfnamefont {L.-C.}\ \bibnamefont {Kwek}}, \bibinfo {author}
  {\bibfnamefont {A.}~\bibnamefont {Minguzzi}},\ and\ \bibinfo {author}
  {\bibfnamefont {W.}~\bibnamefont {von Klitzing}},\ }\bibfield  {title}
  {\bibinfo {title} {Colloquium: {A}tomtronic circuits: {F}rom many-body
  physics to quantum technologies},\ }\href
  {https://doi.org/10.1103%2Frevmodphys.94.041001} {\bibfield  {journal}
  {\bibinfo  {journal} {Reviews of Modern Physics}\ }\textbf {\bibinfo {volume}
  {94}} (\bibinfo {year} {2022})}\BibitemShut {NoStop}%
\bibitem [{\citenamefont {Ryu}\ \emph {et~al.}(2013)\citenamefont {Ryu},
  \citenamefont {Blackburn}, \citenamefont {Blinova},\ and\ \citenamefont
  {Boshier}}]{Ryu_2013}%
  \BibitemOpen
  \bibfield  {author} {\bibinfo {author} {\bibfnamefont {C.}~\bibnamefont
  {Ryu}}, \bibinfo {author} {\bibfnamefont {P.~W.}\ \bibnamefont {Blackburn}},
  \bibinfo {author} {\bibfnamefont {A.~A.}\ \bibnamefont {Blinova}},\ and\
  \bibinfo {author} {\bibfnamefont {M.~G.}\ \bibnamefont {Boshier}},\
  }\bibfield  {title} {\bibinfo {title} {Experimental {R}ealization of
  {J}osephson {J}unctions for an {A}tom {SQUID}},\ }\href
  {https://doi.org/10.1103/PhysRevLett.111.205301} {\bibfield  {journal}
  {\bibinfo  {journal} {Phys. Rev. Lett.}\ }\textbf {\bibinfo {volume} {111}},\
  \bibinfo {pages} {205301} (\bibinfo {year} {2013})}\BibitemShut {NoStop}%
\bibitem [{\citenamefont {Halkyard}\ \emph {et~al.}(2010)\citenamefont
  {Halkyard}, \citenamefont {Jones},\ and\ \citenamefont
  {Gardiner}}]{halkyard2010rotational}%
  \BibitemOpen
  \bibfield  {author} {\bibinfo {author} {\bibfnamefont {P.~L.}\ \bibnamefont
  {Halkyard}}, \bibinfo {author} {\bibfnamefont {M.~P.~A.}\ \bibnamefont
  {Jones}},\ and\ \bibinfo {author} {\bibfnamefont {S.~A.}\ \bibnamefont
  {Gardiner}},\ }\bibfield  {title} {\bibinfo {title} {Rotational response of
  two-component {B}ose-{E}instein condensates in ring traps},\ }\href@noop {}
  {\bibfield  {journal} {\bibinfo  {journal} {Phys. Rev. A}\ }\textbf {\bibinfo
  {volume} {81}},\ \bibinfo {pages} {061602} (\bibinfo {year}
  {2010})}\BibitemShut {NoStop}%
\bibitem [{\citenamefont {Helm}\ \emph {et~al.}(2018)\citenamefont {Helm},
  \citenamefont {Billam}, \citenamefont {Rakonjac}, \citenamefont {Cornish},\
  and\ \citenamefont {Gardiner}}]{Helm_2018}%
  \BibitemOpen
  \bibfield  {author} {\bibinfo {author} {\bibfnamefont {J.~L.}\ \bibnamefont
  {Helm}}, \bibinfo {author} {\bibfnamefont {T.~P.}\ \bibnamefont {Billam}},
  \bibinfo {author} {\bibfnamefont {A.}~\bibnamefont {Rakonjac}}, \bibinfo
  {author} {\bibfnamefont {S.~L.}\ \bibnamefont {Cornish}},\ and\ \bibinfo
  {author} {\bibfnamefont {S.~A.}\ \bibnamefont {Gardiner}},\ }\bibfield
  {title} {\bibinfo {title} {Spin-orbit-coupled interferometry with
  ring-trapped {B}ose-{E}instein condensates},\ }\href@noop {} {\bibfield
  {journal} {\bibinfo  {journal} {Phys. Rev. Lett.}\ }\textbf {\bibinfo
  {volume} {120}},\ \bibinfo {pages} {063201} (\bibinfo {year}
  {2018})}\BibitemShut {NoStop}%
\bibitem [{\citenamefont {Oliinyk}\ \emph {et~al.}(2019)\citenamefont
  {Oliinyk}, \citenamefont {Yakimenko},\ and\ \citenamefont
  {Malomed}}]{Oliinyk_2019}%
  \BibitemOpen
  \bibfield  {author} {\bibinfo {author} {\bibfnamefont {A.}~\bibnamefont
  {Oliinyk}}, \bibinfo {author} {\bibfnamefont {A.}~\bibnamefont {Yakimenko}},\
  and\ \bibinfo {author} {\bibfnamefont {B.}~\bibnamefont {Malomed}},\
  }\bibfield  {title} {\bibinfo {title} {Tunneling of persistent currents in
  coupled ring-shaped {B}ose{\textendash}{E}instein condensates},\ }\href
  {https://doi.org/10.1088/1361-6455/ab46f9} {\bibfield  {journal} {\bibinfo
  {journal} {Journal of Physics B: Atomic, Molecular and Optical Physics}\
  }\textbf {\bibinfo {volume} {52}},\ \bibinfo {pages} {225301} (\bibinfo
  {year} {2019})}\BibitemShut {NoStop}%
\bibitem [{\citenamefont {Oliinyk}\ \emph {et~al.}(2020)\citenamefont
  {Oliinyk}, \citenamefont {Malomed},\ and\ \citenamefont
  {Yakimenko}}]{Oliinyk_2020}%
  \BibitemOpen
  \bibfield  {author} {\bibinfo {author} {\bibfnamefont {A.}~\bibnamefont
  {Oliinyk}}, \bibinfo {author} {\bibfnamefont {B.}~\bibnamefont {Malomed}},\
  and\ \bibinfo {author} {\bibfnamefont {A.}~\bibnamefont {Yakimenko}},\
  }\bibfield  {title} {\bibinfo {title} {Nonlinear dynamics of {J}osephson
  vortices in merging superfluid rings},\ }\href
  {https://doi.org/10.1016/j.cnsns.2019.105113} {\bibfield  {journal} {\bibinfo
   {journal} {Communications in Nonlinear Science and Numerical Simulation}\
  }\textbf {\bibinfo {volume} {83}},\ \bibinfo {pages} {105113} (\bibinfo
  {year} {2020})}\BibitemShut {NoStop}%
\bibitem [{\citenamefont {Kaurov}\ and\ \citenamefont
  {Kuklov}(2005)}]{Kaurov_2005}%
  \BibitemOpen
  \bibfield  {author} {\bibinfo {author} {\bibfnamefont {V.~M.}\ \bibnamefont
  {Kaurov}}\ and\ \bibinfo {author} {\bibfnamefont {A.~B.}\ \bibnamefont
  {Kuklov}},\ }\bibfield  {title} {\bibinfo {title} {Josephson vortex between
  two atomic {B}ose-{E}instein condensates},\ }\href
  {https://doi.org/10.1103/PhysRevA.71.011601} {\bibfield  {journal} {\bibinfo
  {journal} {Phys. Rev. A}\ }\textbf {\bibinfo {volume} {71}},\ \bibinfo
  {pages} {011601} (\bibinfo {year} {2005})}\BibitemShut {NoStop}%
\bibitem [{\citenamefont {Denschlag}\ \emph {et~al.}(2000)\citenamefont
  {Denschlag}, \citenamefont {Simsarian}, \citenamefont {Feder}, \citenamefont
  {Clark}, \citenamefont {Collins}, \citenamefont {Cubizolles}, \citenamefont
  {Deng}, \citenamefont {Hagley}, \citenamefont {Helmerson}, \citenamefont
  {Reinhardt}, \citenamefont {Rolston}, \citenamefont {Schneider},\ and\
  \citenamefont {Phillips}}]{Denschlag_2000}%
  \BibitemOpen
  \bibfield  {author} {\bibinfo {author} {\bibfnamefont {J.}~\bibnamefont
  {Denschlag}}, \bibinfo {author} {\bibfnamefont {J.~E.}\ \bibnamefont
  {Simsarian}}, \bibinfo {author} {\bibfnamefont {D.~L.}\ \bibnamefont
  {Feder}}, \bibinfo {author} {\bibfnamefont {C.~W.}\ \bibnamefont {Clark}},
  \bibinfo {author} {\bibfnamefont {L.~A.}\ \bibnamefont {Collins}}, \bibinfo
  {author} {\bibfnamefont {J.}~\bibnamefont {Cubizolles}}, \bibinfo {author}
  {\bibfnamefont {L.}~\bibnamefont {Deng}}, \bibinfo {author} {\bibfnamefont
  {E.~W.}\ \bibnamefont {Hagley}}, \bibinfo {author} {\bibfnamefont
  {K.}~\bibnamefont {Helmerson}}, \bibinfo {author} {\bibfnamefont {W.~P.}\
  \bibnamefont {Reinhardt}}, \bibinfo {author} {\bibfnamefont {S.~L.}\
  \bibnamefont {Rolston}}, \bibinfo {author} {\bibfnamefont {B.~I.}\
  \bibnamefont {Schneider}},\ and\ \bibinfo {author} {\bibfnamefont {W.~D.}\
  \bibnamefont {Phillips}},\ }\bibfield  {title} {\bibinfo {title} {Generating
  {S}olitons by {P}hase {E}ngineering of a {B}ose-{E}instein {C}ondensate},\
  }\href {https://doi.org/10.1126/science.287.5450.97} {\bibfield  {journal}
  {\bibinfo  {journal} {Science}\ }\textbf {\bibinfo {volume} {287}},\ \bibinfo
  {pages} {97} (\bibinfo {year} {2000})}\BibitemShut {NoStop}%
\bibitem [{\citenamefont {Del~Pace}\ \emph {et~al.}(2022)\citenamefont
  {Del~Pace}, \citenamefont {Xhani}, \citenamefont {Muzi~Falconi},
  \citenamefont {Fedrizzi}, \citenamefont {Grani}, \citenamefont
  {Hernandez~Rajkov}, \citenamefont {Inguscio}, \citenamefont {Scazza},
  \citenamefont {Kwon},\ and\ \citenamefont {Roati}}]{del2022imprinting}%
  \BibitemOpen
  \bibfield  {author} {\bibinfo {author} {\bibfnamefont {G.}~\bibnamefont
  {Del~Pace}}, \bibinfo {author} {\bibfnamefont {K.}~\bibnamefont {Xhani}},
  \bibinfo {author} {\bibfnamefont {A.}~\bibnamefont {Muzi~Falconi}}, \bibinfo
  {author} {\bibfnamefont {M.}~\bibnamefont {Fedrizzi}}, \bibinfo {author}
  {\bibfnamefont {N.}~\bibnamefont {Grani}}, \bibinfo {author} {\bibfnamefont
  {D.}~\bibnamefont {Hernandez~Rajkov}}, \bibinfo {author} {\bibfnamefont
  {M.}~\bibnamefont {Inguscio}}, \bibinfo {author} {\bibfnamefont
  {F.}~\bibnamefont {Scazza}}, \bibinfo {author} {\bibfnamefont {W.~J.}\
  \bibnamefont {Kwon}},\ and\ \bibinfo {author} {\bibfnamefont
  {G.}~\bibnamefont {Roati}},\ }\bibfield  {title} {\bibinfo {title}
  {Imprinting persistent currents in tunable fermionic rings},\ }\href
  {https://doi.org/10.1103/PhysRevX.12.041037} {\bibfield  {journal} {\bibinfo
  {journal} {Phys. Rev. X}\ }\textbf {\bibinfo {volume} {12}},\ \bibinfo
  {pages} {041037} (\bibinfo {year} {2022})}\BibitemShut {NoStop}%
\bibitem [{\citenamefont {Kumar}\ \emph {et~al.}(2018)\citenamefont {Kumar},
  \citenamefont {Dubessy}, \citenamefont {Badr}, \citenamefont {De~Rossi},
  \citenamefont {de~Go\"er~de Herve}, \citenamefont {Longchambon},\ and\
  \citenamefont {Perrin}}]{Kumar_2018}%
  \BibitemOpen
  \bibfield  {author} {\bibinfo {author} {\bibfnamefont {A.}~\bibnamefont
  {Kumar}}, \bibinfo {author} {\bibfnamefont {R.}~\bibnamefont {Dubessy}},
  \bibinfo {author} {\bibfnamefont {T.}~\bibnamefont {Badr}}, \bibinfo {author}
  {\bibfnamefont {C.}~\bibnamefont {De~Rossi}}, \bibinfo {author}
  {\bibfnamefont {M.}~\bibnamefont {de~Go\"er~de Herve}}, \bibinfo {author}
  {\bibfnamefont {L.}~\bibnamefont {Longchambon}},\ and\ \bibinfo {author}
  {\bibfnamefont {H.}~\bibnamefont {Perrin}},\ }\bibfield  {title} {\bibinfo
  {title} {Producing superfluid circulation states using phase imprinting},\
  }\href {https://doi.org/10.1103/PhysRevA.97.043615} {\bibfield  {journal}
  {\bibinfo  {journal} {Phys. Rev. A}\ }\textbf {\bibinfo {volume} {97}},\
  \bibinfo {pages} {043615} (\bibinfo {year} {2018})}\BibitemShut {NoStop}%
\bibitem [{\citenamefont {Allen}\ \emph {et~al.}(1992)\citenamefont {Allen},
  \citenamefont {Beijersbergen}, \citenamefont {Spreeuw},\ and\ \citenamefont
  {Woerdman}}]{Allen_1992}%
  \BibitemOpen
  \bibfield  {author} {\bibinfo {author} {\bibfnamefont {L.}~\bibnamefont
  {Allen}}, \bibinfo {author} {\bibfnamefont {M.~W.}\ \bibnamefont
  {Beijersbergen}}, \bibinfo {author} {\bibfnamefont {R.~J.~C.}\ \bibnamefont
  {Spreeuw}},\ and\ \bibinfo {author} {\bibfnamefont {J.~P.}\ \bibnamefont
  {Woerdman}},\ }\bibfield  {title} {\bibinfo {title} {Orbital angular momentum
  of light and the transformation of {L}aguerre-{G}aussian laser modes},\
  }\href {https://doi.org/10.1103/PhysRevA.45.8185} {\bibfield  {journal}
  {\bibinfo  {journal} {Phys. Rev. A}\ }\textbf {\bibinfo {volume} {45}},\
  \bibinfo {pages} {8185} (\bibinfo {year} {1992})}\BibitemShut {NoStop}%
\bibitem [{\citenamefont {Beattie}\ \emph {et~al.}(2013)\citenamefont
  {Beattie}, \citenamefont {Moulder}, \citenamefont {Fletcher},\ and\
  \citenamefont {Hadzibabic}}]{beattie2013persistent}%
  \BibitemOpen
  \bibfield  {author} {\bibinfo {author} {\bibfnamefont {S.}~\bibnamefont
  {Beattie}}, \bibinfo {author} {\bibfnamefont {S.}~\bibnamefont {Moulder}},
  \bibinfo {author} {\bibfnamefont {R.~J.}\ \bibnamefont {Fletcher}},\ and\
  \bibinfo {author} {\bibfnamefont {Z.}~\bibnamefont {Hadzibabic}},\ }\bibfield
   {title} {\bibinfo {title} {Persistent currents in spinor condensates},\
  }\href@noop {} {\bibfield  {journal} {\bibinfo  {journal} {Phys. Rev. Lett.}\
  }\textbf {\bibinfo {volume} {110}},\ \bibinfo {pages} {025301} (\bibinfo
  {year} {2013})}\BibitemShut {NoStop}%
\bibitem [{\citenamefont {Andersen}\ \emph {et~al.}(2006)\citenamefont
  {Andersen}, \citenamefont {Ryu}, \citenamefont {Clad\'e}, \citenamefont
  {Natarajan}, \citenamefont {Vaziri}, \citenamefont {Helmerson},\ and\
  \citenamefont {Phillips}}]{Andersen_2006}%
  \BibitemOpen
  \bibfield  {author} {\bibinfo {author} {\bibfnamefont {M.~F.}\ \bibnamefont
  {Andersen}}, \bibinfo {author} {\bibfnamefont {C.}~\bibnamefont {Ryu}},
  \bibinfo {author} {\bibfnamefont {P.}~\bibnamefont {Clad\'e}}, \bibinfo
  {author} {\bibfnamefont {V.}~\bibnamefont {Natarajan}}, \bibinfo {author}
  {\bibfnamefont {A.}~\bibnamefont {Vaziri}}, \bibinfo {author} {\bibfnamefont
  {K.}~\bibnamefont {Helmerson}},\ and\ \bibinfo {author} {\bibfnamefont
  {W.~D.}\ \bibnamefont {Phillips}},\ }\bibfield  {title} {\bibinfo {title}
  {Quantized rotation of atoms from photons with orbital angular momentum},\
  }\href {https://doi.org/10.1103/PhysRevLett.97.170406} {\bibfield  {journal}
  {\bibinfo  {journal} {Phys. Rev. Lett.}\ }\textbf {\bibinfo {volume} {97}},\
  \bibinfo {pages} {170406} (\bibinfo {year} {2006})}\BibitemShut {NoStop}%
\bibitem [{\citenamefont {Wright}\ \emph
  {et~al.}(2013{\natexlab{a}})\citenamefont {Wright}, \citenamefont
  {Blakestad}, \citenamefont {Lobb}, \citenamefont {Phillips},\ and\
  \citenamefont {Campbell}}]{wright2013driving}%
  \BibitemOpen
  \bibfield  {author} {\bibinfo {author} {\bibfnamefont {K.~C.}\ \bibnamefont
  {Wright}}, \bibinfo {author} {\bibfnamefont {R.~B.}\ \bibnamefont
  {Blakestad}}, \bibinfo {author} {\bibfnamefont {C.~J.}\ \bibnamefont {Lobb}},
  \bibinfo {author} {\bibfnamefont {W.~D.}\ \bibnamefont {Phillips}},\ and\
  \bibinfo {author} {\bibfnamefont {G.~K.}\ \bibnamefont {Campbell}},\
  }\bibfield  {title} {\bibinfo {title} {Driving phase slips in a superfluid
  atom circuit with a rotating weak link},\ }\href@noop {} {\bibfield
  {journal} {\bibinfo  {journal} {Phys. Rev. Lett.}\ }\textbf {\bibinfo
  {volume} {110}},\ \bibinfo {pages} {025302} (\bibinfo {year}
  {2013}{\natexlab{a}})}\BibitemShut {NoStop}%
\bibitem [{\citenamefont {Eckel}\ \emph
  {et~al.}(2014{\natexlab{a}})\citenamefont {Eckel}, \citenamefont {Lee},
  \citenamefont {Jendrzejewski}, \citenamefont {Murray}, \citenamefont {Clark},
  \citenamefont {Lobb}, \citenamefont {Phillips}, \citenamefont {Edwards},\
  and\ \citenamefont {Campbell}}]{Eckel_2014_Hysteresis}%
  \BibitemOpen
  \bibfield  {author} {\bibinfo {author} {\bibfnamefont {S.}~\bibnamefont
  {Eckel}}, \bibinfo {author} {\bibfnamefont {J.~G.}\ \bibnamefont {Lee}},
  \bibinfo {author} {\bibfnamefont {F.}~\bibnamefont {Jendrzejewski}}, \bibinfo
  {author} {\bibfnamefont {N.}~\bibnamefont {Murray}}, \bibinfo {author}
  {\bibfnamefont {C.~W.}\ \bibnamefont {Clark}}, \bibinfo {author}
  {\bibfnamefont {C.~J.}\ \bibnamefont {Lobb}}, \bibinfo {author}
  {\bibfnamefont {W.~D.}\ \bibnamefont {Phillips}}, \bibinfo {author}
  {\bibfnamefont {M.}~\bibnamefont {Edwards}},\ and\ \bibinfo {author}
  {\bibfnamefont {G.~K.}\ \bibnamefont {Campbell}},\ }\bibfield  {title}
  {\bibinfo {title} {Hysteresis in a quantized superfluid `atomtronic'
  circuit},\ }\href {https://doi.org/10.1038/nature12958} {\bibfield  {journal}
  {\bibinfo  {journal} {Nature}\ }\textbf {\bibinfo {volume} {506}},\ \bibinfo
  {pages} {200} (\bibinfo {year} {2014}{\natexlab{a}})}\BibitemShut {NoStop}%
\bibitem [{\citenamefont {Carleo}\ \emph {et~al.}(2019)\citenamefont {Carleo},
  \citenamefont {Cirac}, \citenamefont {Cranmer}, \citenamefont {Daudet},
  \citenamefont {Schuld}, \citenamefont {Tishby}, \citenamefont
  {Vogt-Maranto},\ and\ \citenamefont {Zdeborov\'a}}]{Carleo_2019}%
  \BibitemOpen
  \bibfield  {author} {\bibinfo {author} {\bibfnamefont {G.}~\bibnamefont
  {Carleo}}, \bibinfo {author} {\bibfnamefont {I.}~\bibnamefont {Cirac}},
  \bibinfo {author} {\bibfnamefont {K.}~\bibnamefont {Cranmer}}, \bibinfo
  {author} {\bibfnamefont {L.}~\bibnamefont {Daudet}}, \bibinfo {author}
  {\bibfnamefont {M.}~\bibnamefont {Schuld}}, \bibinfo {author} {\bibfnamefont
  {N.}~\bibnamefont {Tishby}}, \bibinfo {author} {\bibfnamefont
  {L.}~\bibnamefont {Vogt-Maranto}},\ and\ \bibinfo {author} {\bibfnamefont
  {L.}~\bibnamefont {Zdeborov\'a}},\ }\bibfield  {title} {\bibinfo {title}
  {Machine learning and the physical sciences},\ }\href
  {https://doi.org/10.1103/RevModPhys.91.045002} {\bibfield  {journal}
  {\bibinfo  {journal} {Reviews of Modern Physics}\ }\textbf {\bibinfo {volume}
  {91}},\ \bibinfo {pages} {045002} (\bibinfo {year} {2019})}\BibitemShut
  {NoStop}%
\bibitem [{\citenamefont {Zhou}\ \emph {et~al.}(2018)\citenamefont {Zhou},
  \citenamefont {Tang}, \citenamefont {Liu}, \citenamefont {Pan}, \citenamefont
  {Yan},\ and\ \citenamefont {Zhang}}]{Zhou_2018}%
  \BibitemOpen
  \bibfield  {author} {\bibinfo {author} {\bibfnamefont {Q.}~\bibnamefont
  {Zhou}}, \bibinfo {author} {\bibfnamefont {P.}~\bibnamefont {Tang}}, \bibinfo
  {author} {\bibfnamefont {S.}~\bibnamefont {Liu}}, \bibinfo {author}
  {\bibfnamefont {J.}~\bibnamefont {Pan}}, \bibinfo {author} {\bibfnamefont
  {Q.}~\bibnamefont {Yan}},\ and\ \bibinfo {author} {\bibfnamefont {S.-C.}\
  \bibnamefont {Zhang}},\ }\bibfield  {title} {\bibinfo {title} {Learning atoms
  for materials discovery},\ }\href {https://doi.org/10.1073/pnas.1801181115}
  {\bibfield  {journal} {\bibinfo  {journal} {Proceedings of the National
  Academy of Sciences}\ }\textbf {\bibinfo {volume} {115}},\ \bibinfo {pages}
  {E6411–E6417} (\bibinfo {year} {2018})}\BibitemShut {NoStop}%
\bibitem [{\citenamefont {Barker}\ \emph {et~al.}(2020)\citenamefont {Barker},
  \citenamefont {Style}, \citenamefont {Luksch}, \citenamefont {Sunami},
  \citenamefont {Garrick}, \citenamefont {Hill}, \citenamefont {Foot},\ and\
  \citenamefont {Bentine}}]{Barker_2020}%
  \BibitemOpen
  \bibfield  {author} {\bibinfo {author} {\bibfnamefont {A.~J.}\ \bibnamefont
  {Barker}}, \bibinfo {author} {\bibfnamefont {H.}~\bibnamefont {Style}},
  \bibinfo {author} {\bibfnamefont {K.}~\bibnamefont {Luksch}}, \bibinfo
  {author} {\bibfnamefont {S.}~\bibnamefont {Sunami}}, \bibinfo {author}
  {\bibfnamefont {D.}~\bibnamefont {Garrick}}, \bibinfo {author} {\bibfnamefont
  {F.}~\bibnamefont {Hill}}, \bibinfo {author} {\bibfnamefont {C.~J.}\
  \bibnamefont {Foot}},\ and\ \bibinfo {author} {\bibfnamefont
  {E.}~\bibnamefont {Bentine}},\ }\bibfield  {title} {\bibinfo {title}
  {Applying machine learning optimization methods to the production of a
  quantum gas},\ }\href {https://doi.org/10.1088/2632-2153/ab6432} {\bibfield
  {journal} {\bibinfo  {journal} {Machine Learning: Science and Technology}\
  }\textbf {\bibinfo {volume} {1}},\ \bibinfo {pages} {015007} (\bibinfo {year}
  {2020})}\BibitemShut {NoStop}%
\bibitem [{\citenamefont {Mendelson}\ \emph {et~al.}(2003)\citenamefont
  {Mendelson}, \citenamefont {Smola},\ and\ \citenamefont {van
  Leeuwen}}]{MendelsonShahar2003ALoM}%
  \BibitemOpen
  \bibfield  {author} {\bibinfo {author} {\bibfnamefont {S.}~\bibnamefont
  {Mendelson}}, \bibinfo {author} {\bibfnamefont {A.~J.}\ \bibnamefont
  {Smola}},\ and\ \bibinfo {author} {\bibfnamefont {J.}~\bibnamefont {van
  Leeuwen}},\ }\href@noop {} {\emph {\bibinfo {title} {Advanced Lectures on
  Machine Learning}}},\ \bibinfo {series} {Lecture Notes in Computer Science},
  Vol.\ \bibinfo {volume} {2600}\ (\bibinfo  {publisher} {Springer Berlin /
  Heidelberg},\ \bibinfo {address} {Berlin, Heidelberg},\ \bibinfo {year}
  {2003})\BibitemShut {NoStop}%
\bibitem [{\citenamefont {Jones}\ \emph {et~al.}(1998)\citenamefont {Jones},
  \citenamefont {Schonlau},\ and\ \citenamefont {Welch}}]{Jones_1998}%
  \BibitemOpen
  \bibfield  {author} {\bibinfo {author} {\bibfnamefont {D.}~\bibnamefont
  {Jones}}, \bibinfo {author} {\bibfnamefont {M.}~\bibnamefont {Schonlau}},\
  and\ \bibinfo {author} {\bibfnamefont {W.}~\bibnamefont {Welch}},\ }\bibfield
   {title} {\bibinfo {title} {Efficient {G}lobal {O}ptimization of {E}xpensive
  {B}lack-{B}ox {F}unctions},\ }\href {https://doi.org/10.1023/A:1008306431147}
  {\bibfield  {journal} {\bibinfo  {journal} {Journal of Global Optimization}\
  }\textbf {\bibinfo {volume} {13}},\ \bibinfo {pages} {455} (\bibinfo {year}
  {1998})}\BibitemShut {NoStop}%
\bibitem [{\citenamefont {Bonaccorso}(2018)}]{BonaccorsoGiuseppe2018MMLA}%
  \BibitemOpen
  \bibfield  {author} {\bibinfo {author} {\bibfnamefont {G.}~\bibnamefont
  {Bonaccorso}},\ }\href@noop {} {\emph {\bibinfo {title} {Mastering {M}achine
  {L}earning {A}lgorithms}}},\ \bibinfo {edition} {1st}\ ed.\ (\bibinfo
  {publisher} {Packt Publishing},\ \bibinfo {address} {Birmingham, United
  Kingdom},\ \bibinfo {year} {2018})\BibitemShut {NoStop}%
\bibitem [{\citenamefont {Nakamura}\ \emph {et~al.}(2019)\citenamefont
  {Nakamura}, \citenamefont {Kanemura}, \citenamefont {Nakaso}, \citenamefont
  {Yamamoto},\ and\ \citenamefont {Fukuhara}}]{Nakamura:19}%
  \BibitemOpen
  \bibfield  {author} {\bibinfo {author} {\bibfnamefont {I.}~\bibnamefont
  {Nakamura}}, \bibinfo {author} {\bibfnamefont {A.}~\bibnamefont {Kanemura}},
  \bibinfo {author} {\bibfnamefont {T.}~\bibnamefont {Nakaso}}, \bibinfo
  {author} {\bibfnamefont {R.}~\bibnamefont {Yamamoto}},\ and\ \bibinfo
  {author} {\bibfnamefont {T.}~\bibnamefont {Fukuhara}},\ }\bibfield  {title}
  {\bibinfo {title} {Non-standard trajectories found by machine learning for
  evaporative cooling of $^{87}${R}b atoms},\ }\href
  {https://doi.org/10.1364/OE.27.020435} {\bibfield  {journal} {\bibinfo
  {journal} {Optics Express}\ }\textbf {\bibinfo {volume} {27}},\ \bibinfo
  {pages} {20435} (\bibinfo {year} {2019})}\BibitemShut {NoStop}%
\bibitem [{\citenamefont {Bell}\ \emph {et~al.}(2016)\citenamefont {Bell},
  \citenamefont {Glidden}, \citenamefont {Humbert}, \citenamefont {Bromley},
  \citenamefont {Haine}, \citenamefont {Davis}, \citenamefont {Neely},
  \citenamefont {Baker},\ and\ \citenamefont {Rubinsztein-Dunlop}}]{Bell_2016}%
  \BibitemOpen
  \bibfield  {author} {\bibinfo {author} {\bibfnamefont {T.~A.}\ \bibnamefont
  {Bell}}, \bibinfo {author} {\bibfnamefont {J.~A.~P.}\ \bibnamefont
  {Glidden}}, \bibinfo {author} {\bibfnamefont {L.}~\bibnamefont {Humbert}},
  \bibinfo {author} {\bibfnamefont {M.~W.~J.}\ \bibnamefont {Bromley}},
  \bibinfo {author} {\bibfnamefont {S.~A.}\ \bibnamefont {Haine}}, \bibinfo
  {author} {\bibfnamefont {M.~J.}\ \bibnamefont {Davis}}, \bibinfo {author}
  {\bibfnamefont {T.~W.}\ \bibnamefont {Neely}}, \bibinfo {author}
  {\bibfnamefont {M.~A.}\ \bibnamefont {Baker}},\ and\ \bibinfo {author}
  {\bibfnamefont {H.}~\bibnamefont {Rubinsztein-Dunlop}},\ }\bibfield  {title}
  {\bibinfo {title} {Bose-{E}instein condensation in large time-averaged
  optical ring potentials},\ }\href
  {https://doi.org/10.1088/1367-2630/18/3/035003} {\bibfield  {journal}
  {\bibinfo  {journal} {New Journal of Physics}\ }\textbf {\bibinfo {volume}
  {18}},\ \bibinfo {pages} {035003} (\bibinfo {year} {2016})}\BibitemShut
  {NoStop}%
\bibitem [{\citenamefont {Gauthier}(2019)}]{Gauthier}%
  \BibitemOpen
  \bibfield  {author} {\bibinfo {author} {\bibfnamefont {G.}~\bibnamefont
  {Gauthier}},\ }\emph {\bibinfo {title} {Transport and turbulence in
  quasi-uniform and versatile {B}ose-{E}instein condensates}},\ \href
  {https://doi.org/10.14264/uql.2019.682} {Ph.D. thesis},\ \bibinfo  {school}
  {The University of Queensland} (\bibinfo {year} {2019})\BibitemShut {NoStop}%
\bibitem [{\citenamefont {Wright}\ \emph
  {et~al.}(2013{\natexlab{b}})\citenamefont {Wright}, \citenamefont
  {Blakestad}, \citenamefont {Lobb}, \citenamefont {Phillips},\ and\
  \citenamefont {Campbell}}]{Wright_2013}%
  \BibitemOpen
  \bibfield  {author} {\bibinfo {author} {\bibfnamefont {K.~C.}\ \bibnamefont
  {Wright}}, \bibinfo {author} {\bibfnamefont {R.~B.}\ \bibnamefont
  {Blakestad}}, \bibinfo {author} {\bibfnamefont {C.~J.}\ \bibnamefont {Lobb}},
  \bibinfo {author} {\bibfnamefont {W.~D.}\ \bibnamefont {Phillips}},\ and\
  \bibinfo {author} {\bibfnamefont {G.~K.}\ \bibnamefont {Campbell}},\
  }\bibfield  {title} {\bibinfo {title} {Threshold for creating excitations in
  a stirred superfluid ring},\ }\href
  {https://doi.org/10.1103/PhysRevA.88.063633} {\bibfield  {journal} {\bibinfo
  {journal} {Phys. Rev. A}\ }\textbf {\bibinfo {volume} {88}},\ \bibinfo
  {pages} {063633} (\bibinfo {year} {2013}{\natexlab{b}})}\BibitemShut
  {NoStop}%
\bibitem [{\citenamefont {Yakimenko}\ \emph {et~al.}(2015)\citenamefont
  {Yakimenko}, \citenamefont {Isaieva}, \citenamefont {Vilchinskii},\ and\
  \citenamefont {Ostrovskaya}}]{Yakimenko_2015_VE}%
  \BibitemOpen
  \bibfield  {author} {\bibinfo {author} {\bibfnamefont {A.~I.}\ \bibnamefont
  {Yakimenko}}, \bibinfo {author} {\bibfnamefont {K.~O.}\ \bibnamefont
  {Isaieva}}, \bibinfo {author} {\bibfnamefont {S.~I.}\ \bibnamefont
  {Vilchinskii}},\ and\ \bibinfo {author} {\bibfnamefont {E.~A.}\ \bibnamefont
  {Ostrovskaya}},\ }\bibfield  {title} {\bibinfo {title} {Vortex excitation in
  a stirred toroidal {B}ose-{E}instein condensate},\ }\href
  {https://doi.org/10.1103/PhysRevA.91.023607} {\bibfield  {journal} {\bibinfo
  {journal} {Phys. Rev. A}\ }\textbf {\bibinfo {volume} {91}},\ \bibinfo
  {pages} {023607} (\bibinfo {year} {2015})}\BibitemShut {NoStop}%
\bibitem [{\citenamefont {Murray}\ \emph {et~al.}(2013)\citenamefont {Murray},
  \citenamefont {Krygier}, \citenamefont {Edwards}, \citenamefont {Wright},
  \citenamefont {Campbell},\ and\ \citenamefont {Clark}}]{Murray_2013}%
  \BibitemOpen
  \bibfield  {author} {\bibinfo {author} {\bibfnamefont {N.}~\bibnamefont
  {Murray}}, \bibinfo {author} {\bibfnamefont {M.}~\bibnamefont {Krygier}},
  \bibinfo {author} {\bibfnamefont {M.}~\bibnamefont {Edwards}}, \bibinfo
  {author} {\bibfnamefont {K.~C.}\ \bibnamefont {Wright}}, \bibinfo {author}
  {\bibfnamefont {G.~K.}\ \bibnamefont {Campbell}},\ and\ \bibinfo {author}
  {\bibfnamefont {C.~W.}\ \bibnamefont {Clark}},\ }\bibfield  {title} {\bibinfo
  {title} {Probing the circulation of ring-shaped {B}ose-{E}instein
  condensates},\ }\href {https://doi.org/10.1103/PhysRevA.88.053615} {\bibfield
   {journal} {\bibinfo  {journal} {Phys. Rev. A}\ }\textbf {\bibinfo {volume}
  {88}},\ \bibinfo {pages} {053615} (\bibinfo {year} {2013})}\BibitemShut
  {NoStop}%
\bibitem [{\citenamefont {Haug}\ \emph {et~al.}(2021)\citenamefont {Haug},
  \citenamefont {Dumke}, \citenamefont {Kwek}, \citenamefont {Miniatura},\ and\
  \citenamefont {Amico}}]{Amico_2021}%
  \BibitemOpen
  \bibfield  {author} {\bibinfo {author} {\bibfnamefont {T.}~\bibnamefont
  {Haug}}, \bibinfo {author} {\bibfnamefont {R.}~\bibnamefont {Dumke}},
  \bibinfo {author} {\bibfnamefont {L.-C.}\ \bibnamefont {Kwek}}, \bibinfo
  {author} {\bibfnamefont {C.}~\bibnamefont {Miniatura}},\ and\ \bibinfo
  {author} {\bibfnamefont {L.}~\bibnamefont {Amico}},\ }\bibfield  {title}
  {\bibinfo {title} {Machine-learning engineering of quantum currents},\ }\href
  {https://doi.org/10.1103/PhysRevResearch.3.013034} {\bibfield  {journal}
  {\bibinfo  {journal} {Phys. Rev. Research}\ }\textbf {\bibinfo {volume}
  {3}},\ \bibinfo {pages} {013034} (\bibinfo {year} {2021})}\BibitemShut
  {NoStop}%
\bibitem [{\citenamefont {Wigley}\ \emph {et~al.}(2016)\citenamefont {Wigley},
  \citenamefont {Everitt}, \citenamefont {van~den Hengel}, \citenamefont
  {Bastian}, \citenamefont {Sooriyabandara}, \citenamefont {McDonald},
  \citenamefont {Hardman}, \citenamefont {Quinlivan}, \citenamefont {Manju},\
  and\ \citenamefont {Kuhn}}]{Wigley_2016}%
  \BibitemOpen
  \bibfield  {author} {\bibinfo {author} {\bibfnamefont {P.~B.}\ \bibnamefont
  {Wigley}}, \bibinfo {author} {\bibfnamefont {P.~J.}\ \bibnamefont {Everitt}},
  \bibinfo {author} {\bibfnamefont {A.}~\bibnamefont {van~den Hengel}},
  \bibinfo {author} {\bibfnamefont {J.~W.}\ \bibnamefont {Bastian}}, \bibinfo
  {author} {\bibfnamefont {M.~A.}\ \bibnamefont {Sooriyabandara}}, \bibinfo
  {author} {\bibfnamefont {G.~D.}\ \bibnamefont {McDonald}}, \bibinfo {author}
  {\bibfnamefont {K.~S.}\ \bibnamefont {Hardman}}, \bibinfo {author}
  {\bibfnamefont {C.~D.}\ \bibnamefont {Quinlivan}}, \bibinfo {author}
  {\bibfnamefont {P.}~\bibnamefont {Manju}},\ and\ \bibinfo {author}
  {\bibfnamefont {C.~C.~N.}\ \bibnamefont {Kuhn}},\ }\bibfield  {title}
  {\bibinfo {title} {Fast machine-learning online optimization of
  ultra-cold-atom experiments},\ }\href@noop {} {\bibfield  {journal} {\bibinfo
   {journal} {Scientific Reports}\ }\textbf {\bibinfo {volume} {6}} (\bibinfo
  {year} {2016})}\BibitemShut {NoStop}%
\bibitem [{\citenamefont {Gauthier}\ \emph {et~al.}(2016)\citenamefont
  {Gauthier}, \citenamefont {Lenton}, \citenamefont {Parry}, \citenamefont
  {Baker}, \citenamefont {Davis}, \citenamefont {Rubinsztein-Dunlop},\ and\
  \citenamefont {Neely}}]{Gauthier:16}%
  \BibitemOpen
  \bibfield  {author} {\bibinfo {author} {\bibfnamefont {G.}~\bibnamefont
  {Gauthier}}, \bibinfo {author} {\bibfnamefont {I.}~\bibnamefont {Lenton}},
  \bibinfo {author} {\bibfnamefont {N.~M.}\ \bibnamefont {Parry}}, \bibinfo
  {author} {\bibfnamefont {M.}~\bibnamefont {Baker}}, \bibinfo {author}
  {\bibfnamefont {M.~J.}\ \bibnamefont {Davis}}, \bibinfo {author}
  {\bibfnamefont {H.}~\bibnamefont {Rubinsztein-Dunlop}},\ and\ \bibinfo
  {author} {\bibfnamefont {T.~W.}\ \bibnamefont {Neely}},\ }\bibfield  {title}
  {\bibinfo {title} {Direct imaging of a digital-micromirror device for
  configurable microscopic optical potentials},\ }\href
  {https://doi.org/10.1364/OPTICA.3.001136} {\bibfield  {journal} {\bibinfo
  {journal} {Optica}\ }\textbf {\bibinfo {volume} {3}},\ \bibinfo {pages}
  {1136} (\bibinfo {year} {2016})}\BibitemShut {NoStop}%
\bibitem [{\citenamefont {Gauthier}\ \emph {et~al.}(2020)\citenamefont
  {Gauthier}, \citenamefont {Bell}, \citenamefont {Baker}, \citenamefont
  {Rubinsztein-Dunlop},\ and\ \citenamefont {Neely}}]{Gauthier_2020}%
  \BibitemOpen
  \bibfield  {author} {\bibinfo {author} {\bibfnamefont {G.}~\bibnamefont
  {Gauthier}}, \bibinfo {author} {\bibfnamefont {T.~A.}\ \bibnamefont {Bell}},
  \bibinfo {author} {\bibfnamefont {M.}~\bibnamefont {Baker}}, \bibinfo
  {author} {\bibfnamefont {H.}~\bibnamefont {Rubinsztein-Dunlop}},\ and\
  \bibinfo {author} {\bibfnamefont {T.~W.}\ \bibnamefont {Neely}},\ }\bibfield
  {title} {\bibinfo {title} {Feedforward {O}ptimisation of {O}ptical {T}rapping
  {P}otentials for {U}ltracold {A}toms},\ }in\ \href
  {https://doi.org/10.1364/CLEOPR.2020.C8C_4} {\emph {\bibinfo {booktitle}
  {2020 Conference on Lasers and Electro-Optics Pacific Rim (CLEO-PR)}}}\
  (\bibinfo {year} {2020})\ pp.\ \bibinfo {pages} {1--2}\BibitemShut {NoStop}%
\bibitem [{\citenamefont {Andrews}\ \emph {et~al.}(1997)\citenamefont
  {Andrews}, \citenamefont {Townsend}, \citenamefont {Miesner}, \citenamefont
  {Durfee}, \citenamefont {Kurn},\ and\ \citenamefont
  {Ketterle}}]{Andrews_1997}%
  \BibitemOpen
  \bibfield  {author} {\bibinfo {author} {\bibfnamefont {M.~R.}\ \bibnamefont
  {Andrews}}, \bibinfo {author} {\bibfnamefont {C.~G.}\ \bibnamefont
  {Townsend}}, \bibinfo {author} {\bibfnamefont {H.-J.}\ \bibnamefont
  {Miesner}}, \bibinfo {author} {\bibfnamefont {D.~S.}\ \bibnamefont {Durfee}},
  \bibinfo {author} {\bibfnamefont {D.~M.}\ \bibnamefont {Kurn}},\ and\
  \bibinfo {author} {\bibfnamefont {W.}~\bibnamefont {Ketterle}},\ }\bibfield
  {title} {\bibinfo {title} {Observation of {I}nterference {B}etween {T}wo
  {B}ose {C}ondensates},\ }\href {https://doi.org/10.1126/science.275.5300.637}
  {\bibfield  {journal} {\bibinfo  {journal} {Science}\ }\textbf {\bibinfo
  {volume} {275}},\ \bibinfo {pages} {637} (\bibinfo {year}
  {1997})}\BibitemShut {NoStop}%
\bibitem [{\citenamefont {Eckel}\ \emph
  {et~al.}(2014{\natexlab{b}})\citenamefont {Eckel}, \citenamefont
  {Jendrzejewski}, \citenamefont {Kumar}, \citenamefont {Lobb},\ and\
  \citenamefont {Campbell}}]{Eckel_2014}%
  \BibitemOpen
  \bibfield  {author} {\bibinfo {author} {\bibfnamefont {S.}~\bibnamefont
  {Eckel}}, \bibinfo {author} {\bibfnamefont {F.}~\bibnamefont
  {Jendrzejewski}}, \bibinfo {author} {\bibfnamefont {A.}~\bibnamefont
  {Kumar}}, \bibinfo {author} {\bibfnamefont {C.~J.}\ \bibnamefont {Lobb}},\
  and\ \bibinfo {author} {\bibfnamefont {G.~K.}\ \bibnamefont {Campbell}},\
  }\bibfield  {title} {\bibinfo {title} {Interferometric {M}easurement of the
  {C}urrent-{P}hase {R}elationship of a {S}uperfluid {W}eak {L}ink},\ }\href
  {https://doi.org/10.1103/PhysRevX.4.031052} {\bibfield  {journal} {\bibinfo
  {journal} {Phys. Rev. X}\ }\textbf {\bibinfo {volume} {4}},\ \bibinfo {pages}
  {031052} (\bibinfo {year} {2014}{\natexlab{b}})}\BibitemShut {NoStop}%
\bibitem [{\citenamefont {Wigley}(2017)}]{Wigley_Thesis}%
  \BibitemOpen
  \bibfield  {author} {\bibinfo {author} {\bibfnamefont {P.~B.}\ \bibnamefont
  {Wigley}},\ }\emph {\bibinfo {title} {Generating and observing soliton
  dynamics in {B}ose-{E}instein {C}ondensates}},\ \href@noop {} {Ph.D.
  thesis},\ \bibinfo  {school} {Department of Quantum Science, Research School
  of Physics and Engineering, The Australian National University} (\bibinfo
  {year} {2017})\BibitemShut {NoStop}%
\bibitem [{\citenamefont {Mehta}\ \emph {et~al.}(2019)\citenamefont {Mehta},
  \citenamefont {Bukov}, \citenamefont {Wang}, \citenamefont {Day},
  \citenamefont {Richardson}, \citenamefont {Fisher},\ and\ \citenamefont
  {Schwab}}]{MEHTA20191}%
  \BibitemOpen
  \bibfield  {author} {\bibinfo {author} {\bibfnamefont {P.}~\bibnamefont
  {Mehta}}, \bibinfo {author} {\bibfnamefont {M.}~\bibnamefont {Bukov}},
  \bibinfo {author} {\bibfnamefont {C.-H.}\ \bibnamefont {Wang}}, \bibinfo
  {author} {\bibfnamefont {A.~G.}\ \bibnamefont {Day}}, \bibinfo {author}
  {\bibfnamefont {C.}~\bibnamefont {Richardson}}, \bibinfo {author}
  {\bibfnamefont {C.~K.}\ \bibnamefont {Fisher}},\ and\ \bibinfo {author}
  {\bibfnamefont {D.~J.}\ \bibnamefont {Schwab}},\ }\bibfield  {title}
  {\bibinfo {title} {A high-bias, low-variance introduction to machine learning
  for physicists},\ }\href
  {https://doi.org/https://doi.org/10.1016/j.physrep.2019.03.001} {\bibfield
  {journal} {\bibinfo  {journal} {Physics Reports}\ }\textbf {\bibinfo {volume}
  {810}},\ \bibinfo {pages} {1} (\bibinfo {year} {2019})}\BibitemShut {NoStop}%
\bibitem [{Note1()}]{Note1}%
  \BibitemOpen
  \bibinfo {note} {With very few fringes present, the fringes themselves wrap
  around the entire interference region. The detection algorithm effectively
  counts periodicity of density in these regions meaning that such wrapping
  obscures the measurement.}\BibitemShut {Stop}%
\bibitem [{\citenamefont {Kiehn}\ \emph {et~al.}(2022)\citenamefont {Kiehn},
  \citenamefont {Singh},\ and\ \citenamefont {Mathey}}]{Kiehn_2022}%
  \BibitemOpen
  \bibfield  {author} {\bibinfo {author} {\bibfnamefont {H.}~\bibnamefont
  {Kiehn}}, \bibinfo {author} {\bibfnamefont {V.~P.}\ \bibnamefont {Singh}},\
  and\ \bibinfo {author} {\bibfnamefont {L.}~\bibnamefont {Mathey}},\
  }\bibfield  {title} {\bibinfo {title} {Superfluidity of a laser-stirred
  {B}ose-{E}instein condensate},\ }\href
  {https://doi.org/10.1103/PhysRevA.105.043317} {\bibfield  {journal} {\bibinfo
   {journal} {Phys. Rev. A}\ }\textbf {\bibinfo {volume} {105}},\ \bibinfo
  {pages} {043317} (\bibinfo {year} {2022})}\BibitemShut {NoStop}%
\bibitem [{\citenamefont {Baggaley}\ and\ \citenamefont
  {Parker}(2018)}]{Baggaley_2018}%
  \BibitemOpen
  \bibfield  {author} {\bibinfo {author} {\bibfnamefont {A.~W.}\ \bibnamefont
  {Baggaley}}\ and\ \bibinfo {author} {\bibfnamefont {N.~G.}\ \bibnamefont
  {Parker}},\ }\bibfield  {title} {\bibinfo {title} {Kelvin-{H}elmholtz
  instability in a single-component atomic superfluid},\ }\href@noop {}
  {\bibfield  {journal} {\bibinfo  {journal} {Phys. Rev. A}\ }\textbf {\bibinfo
  {volume} {97}},\ \bibinfo {pages} {053608} (\bibinfo {year}
  {2018})}\BibitemShut {NoStop}%
\bibitem [{\citenamefont {Hernandez-Rajkov}\ \emph {et~al.}(2023)\citenamefont
  {Hernandez-Rajkov}, \citenamefont {Grani}, \citenamefont {Scazza},
  \citenamefont {Del~Pace}, \citenamefont {Kwon}, \citenamefont {Inguscio},
  \citenamefont {Xhani}, \citenamefont {Fort}, \citenamefont {Modugno},
  \citenamefont {Marino} \emph {et~al.}}]{hernandez2023universality}%
  \BibitemOpen
  \bibfield  {author} {\bibinfo {author} {\bibfnamefont {D.}~\bibnamefont
  {Hernandez-Rajkov}}, \bibinfo {author} {\bibfnamefont {N.}~\bibnamefont
  {Grani}}, \bibinfo {author} {\bibfnamefont {F.}~\bibnamefont {Scazza}},
  \bibinfo {author} {\bibfnamefont {G.}~\bibnamefont {Del~Pace}}, \bibinfo
  {author} {\bibfnamefont {W.}~\bibnamefont {Kwon}}, \bibinfo {author}
  {\bibfnamefont {M.}~\bibnamefont {Inguscio}}, \bibinfo {author}
  {\bibfnamefont {K.}~\bibnamefont {Xhani}}, \bibinfo {author} {\bibfnamefont
  {C.}~\bibnamefont {Fort}}, \bibinfo {author} {\bibfnamefont {M.}~\bibnamefont
  {Modugno}}, \bibinfo {author} {\bibfnamefont {F.}~\bibnamefont {Marino}},
  \emph {et~al.},\ }\bibfield  {title} {\bibinfo {title} {Universality of the
  superfluid {K}elvin-{H}elmholtz instability by single-vortex tracking},\
  }\href@noop {} {\bibfield  {journal} {\bibinfo  {journal} {arXiv preprint
  arXiv:2303.12631}\ } (\bibinfo {year} {2023})}\BibitemShut {NoStop}%
\bibitem [{\citenamefont {Reeves}\ \emph {et~al.}(2015)\citenamefont {Reeves},
  \citenamefont {Billam}, \citenamefont {Anderson},\ and\ \citenamefont
  {Bradley}}]{Reeves_2015}%
  \BibitemOpen
  \bibfield  {author} {\bibinfo {author} {\bibfnamefont {M.~T.}\ \bibnamefont
  {Reeves}}, \bibinfo {author} {\bibfnamefont {T.~P.}\ \bibnamefont {Billam}},
  \bibinfo {author} {\bibfnamefont {B.~P.}\ \bibnamefont {Anderson}},\ and\
  \bibinfo {author} {\bibfnamefont {A.~S.}\ \bibnamefont {Bradley}},\
  }\bibfield  {title} {\bibinfo {title} {Identifying a {S}uperfluid {R}eynolds
  {N}umber via {D}ynamical {S}imilarity},\ }\href
  {https://doi.org/10.1103/PhysRevLett.114.155302} {\bibfield  {journal}
  {\bibinfo  {journal} {Phys. Rev. Lett.}\ }\textbf {\bibinfo {volume} {114}},\
  \bibinfo {pages} {155302} (\bibinfo {year} {2015})}\BibitemShut {NoStop}%
\bibitem [{\citenamefont {Raman}\ \emph {et~al.}(1999)\citenamefont {Raman},
  \citenamefont {K{\"o}hl}, \citenamefont {Onofrio}, \citenamefont {Durfee},
  \citenamefont {Kuklewicz}, \citenamefont {Hadzibabic},\ and\ \citenamefont
  {Ketterle}}]{raman1999evidence}%
  \BibitemOpen
  \bibfield  {author} {\bibinfo {author} {\bibfnamefont {C.}~\bibnamefont
  {Raman}}, \bibinfo {author} {\bibfnamefont {M.}~\bibnamefont {K{\"o}hl}},
  \bibinfo {author} {\bibfnamefont {R.}~\bibnamefont {Onofrio}}, \bibinfo
  {author} {\bibfnamefont {D.~S.}\ \bibnamefont {Durfee}}, \bibinfo {author}
  {\bibfnamefont {C.~E.}\ \bibnamefont {Kuklewicz}}, \bibinfo {author}
  {\bibfnamefont {Z.}~\bibnamefont {Hadzibabic}},\ and\ \bibinfo {author}
  {\bibfnamefont {W.}~\bibnamefont {Ketterle}},\ }\bibfield  {title} {\bibinfo
  {title} {Evidence for a critical velocity in a {B}ose-{E}instein condensed
  gas},\ }\href@noop {} {\bibfield  {journal} {\bibinfo  {journal} {Phys. Rev.
  Lett.}\ }\textbf {\bibinfo {volume} {83}},\ \bibinfo {pages} {2502} (\bibinfo
  {year} {1999})}\BibitemShut {NoStop}%
\bibitem [{\citenamefont {Stie\ss{}berger}\ and\ \citenamefont
  {Zwerger}(2000)}]{Stiesberger_2000}%
  \BibitemOpen
  \bibfield  {author} {\bibinfo {author} {\bibfnamefont {J.~S.}\ \bibnamefont
  {Stie\ss{}berger}}\ and\ \bibinfo {author} {\bibfnamefont {W.}~\bibnamefont
  {Zwerger}},\ }\bibfield  {title} {\bibinfo {title} {Critcal velocity of
  superfluid flow past large obstacles in {B}ose-{E}instein condensates},\
  }\href {https://doi.org/10.1103/PhysRevA.62.061601} {\bibfield  {journal}
  {\bibinfo  {journal} {Phys. Rev. A}\ }\textbf {\bibinfo {volume} {62}},\
  \bibinfo {pages} {061601} (\bibinfo {year} {2000})}\BibitemShut {NoStop}%
\bibitem [{\citenamefont {Kwon}\ \emph {et~al.}(2015)\citenamefont {Kwon},
  \citenamefont {Moon}, \citenamefont {Seo},\ and\ \citenamefont
  {Shin}}]{Kwon_2015}%
  \BibitemOpen
  \bibfield  {author} {\bibinfo {author} {\bibfnamefont {W.}~\bibnamefont
  {Kwon}}, \bibinfo {author} {\bibfnamefont {G.}~\bibnamefont {Moon}}, \bibinfo
  {author} {\bibfnamefont {S.}~\bibnamefont {Seo}},\ and\ \bibinfo {author}
  {\bibfnamefont {Y.-i.}\ \bibnamefont {Shin}},\ }\bibfield  {title} {\bibinfo
  {title} {Critical {V}elocity for {V}ortex {S}hedding in a {B}ose-{E}instein
  {C}ondensate},\ }\href {https://doi.org/10.1103/PhysRevA.91.053615}
  {\bibfield  {journal} {\bibinfo  {journal} {Phys. Rev. A}\ }\textbf {\bibinfo
  {volume} {91}} (\bibinfo {year} {2015})}\BibitemShut {NoStop}%
\bibitem [{Note2()}]{Note2}%
  \BibitemOpen
  \bibinfo {note} {We note that the lower winding numbers in O.III and O.IV
  result from the strong weighting on the stirring time punishment term in the
  cost functions. This observation highlights the trade-offs introduced in the
  cost functions which are likely the limiting factors preventing more complete
  optimization in more complicated scenarios.}\BibitemShut {Stop}%
\bibitem [{\citenamefont {Menchon-Enrich}\ \emph {et~al.}(2016)\citenamefont
  {Menchon-Enrich}, \citenamefont {Benseny}, \citenamefont {Ahufinger},
  \citenamefont {Greentree}, \citenamefont {Busch},\ and\ \citenamefont
  {Mompart}}]{Menchon_2016}%
  \BibitemOpen
  \bibfield  {author} {\bibinfo {author} {\bibfnamefont {R.}~\bibnamefont
  {Menchon-Enrich}}, \bibinfo {author} {\bibfnamefont {A.}~\bibnamefont
  {Benseny}}, \bibinfo {author} {\bibfnamefont {V.}~\bibnamefont {Ahufinger}},
  \bibinfo {author} {\bibfnamefont {A.~D.}\ \bibnamefont {Greentree}}, \bibinfo
  {author} {\bibfnamefont {T.}~\bibnamefont {Busch}},\ and\ \bibinfo {author}
  {\bibfnamefont {J.}~\bibnamefont {Mompart}},\ }\bibfield  {title} {\bibinfo
  {title} {Spatial adiabatic passage: a review of recent progress},\
  }\href@noop {} {\bibfield  {journal} {\bibinfo  {journal} {Reports on
  Progress in Physics}\ }\textbf {\bibinfo {volume} {79}},\ \bibinfo {pages}
  {074401} (\bibinfo {year} {2016})}\BibitemShut {NoStop}%
\bibitem [{\citenamefont {Gu{\'e}ry-Odelin}\ \emph {et~al.}(2019)\citenamefont
  {Gu{\'e}ry-Odelin}, \citenamefont {Ruschhaupt}, \citenamefont {Kiely},
  \citenamefont {Torrontegui}, \citenamefont {Mart{\'\i}nez-Garaot},\ and\
  \citenamefont {Muga}}]{Guery_2019}%
  \BibitemOpen
  \bibfield  {author} {\bibinfo {author} {\bibfnamefont {D.}~\bibnamefont
  {Gu{\'e}ry-Odelin}}, \bibinfo {author} {\bibfnamefont {A.}~\bibnamefont
  {Ruschhaupt}}, \bibinfo {author} {\bibfnamefont {A.}~\bibnamefont {Kiely}},
  \bibinfo {author} {\bibfnamefont {E.}~\bibnamefont {Torrontegui}}, \bibinfo
  {author} {\bibfnamefont {S.}~\bibnamefont {Mart{\'\i}nez-Garaot}},\ and\
  \bibinfo {author} {\bibfnamefont {J.~G.}\ \bibnamefont {Muga}},\ }\bibfield
  {title} {\bibinfo {title} {Shortcuts to adiabaticity: {C}oncepts, methods,
  and applications},\ }\href@noop {} {\bibfield  {journal} {\bibinfo  {journal}
  {Reviews of Modern Physics}\ }\textbf {\bibinfo {volume} {91}},\ \bibinfo
  {pages} {045001} (\bibinfo {year} {2019})}\BibitemShut {NoStop}%
\bibitem [{\citenamefont {Pandey}\ \emph {et~al.}(2019)\citenamefont {Pandey},
  \citenamefont {Mas}, \citenamefont {Drougakis}, \citenamefont {Thekkeppatt},
  \citenamefont {Bolpasi}, \citenamefont {Vasilakis}, \citenamefont {Poulios},\
  and\ \citenamefont {von Klitzing}}]{Pandey_2019}%
  \BibitemOpen
  \bibfield  {author} {\bibinfo {author} {\bibfnamefont {S.}~\bibnamefont
  {Pandey}}, \bibinfo {author} {\bibfnamefont {H.}~\bibnamefont {Mas}},
  \bibinfo {author} {\bibfnamefont {G.}~\bibnamefont {Drougakis}}, \bibinfo
  {author} {\bibfnamefont {P.}~\bibnamefont {Thekkeppatt}}, \bibinfo {author}
  {\bibfnamefont {V.}~\bibnamefont {Bolpasi}}, \bibinfo {author} {\bibfnamefont
  {G.}~\bibnamefont {Vasilakis}}, \bibinfo {author} {\bibfnamefont
  {K.}~\bibnamefont {Poulios}},\ and\ \bibinfo {author} {\bibfnamefont
  {W.}~\bibnamefont {von Klitzing}},\ }\bibfield  {title} {\bibinfo {title}
  {Hypersonic {B}ose{\textendash}{E}instein condensates in accelerator rings},\
  }\href {https://doi.org/10.1038/s41586-019-1273-5} {\bibfield  {journal}
  {\bibinfo  {journal} {Nature}\ }\textbf {\bibinfo {volume} {570}},\ \bibinfo
  {pages} {205} (\bibinfo {year} {2019})}\BibitemShut {NoStop}%
\bibitem [{\citenamefont {Edwards}(2013)}]{Edwards_2013}%
  \BibitemOpen
  \bibfield  {author} {\bibinfo {author} {\bibfnamefont {M.}~\bibnamefont
  {Edwards}},\ }\bibfield  {title} {\bibinfo {title} {Atom {S}{Q}{U}{I}{D}},\
  }\href {https://doi.org/10.1038/nphys2546} {\bibfield  {journal} {\bibinfo
  {journal} {Nature Physics}\ }\textbf {\bibinfo {volume} {9}},\ \bibinfo
  {pages} {68} (\bibinfo {year} {2013})}\BibitemShut {NoStop}%
\bibitem [{\citenamefont {Gross}\ \emph {et~al.}(2010)\citenamefont {Gross},
  \citenamefont {Zibold}, \citenamefont {Nicklas}, \citenamefont {Est{\`e}ve},\
  and\ \citenamefont {Oberthaler}}]{Gross2010}%
  \BibitemOpen
  \bibfield  {author} {\bibinfo {author} {\bibfnamefont {C.}~\bibnamefont
  {Gross}}, \bibinfo {author} {\bibfnamefont {T.}~\bibnamefont {Zibold}},
  \bibinfo {author} {\bibfnamefont {E.}~\bibnamefont {Nicklas}}, \bibinfo
  {author} {\bibfnamefont {J.}~\bibnamefont {Est{\`e}ve}},\ and\ \bibinfo
  {author} {\bibfnamefont {M.~K.}\ \bibnamefont {Oberthaler}},\ }\bibfield
  {title} {\bibinfo {title} {Nonlinear atom interferometer surpasses classical
  precision limit},\ }\href {https://doi.org/10.1038/nature08919} {\bibfield
  {journal} {\bibinfo  {journal} {Nature}\ }\textbf {\bibinfo {volume} {464}},\
  \bibinfo {pages} {1165} (\bibinfo {year} {2010})}\BibitemShut {NoStop}%
\bibitem [{\citenamefont {Levy}\ \emph {et~al.}(2007)\citenamefont {Levy},
  \citenamefont {Lahoud}, \citenamefont {Shomroni},\ and\ \citenamefont
  {Steinhauer}}]{Levy2007}%
  \BibitemOpen
  \bibfield  {author} {\bibinfo {author} {\bibfnamefont {S.}~\bibnamefont
  {Levy}}, \bibinfo {author} {\bibfnamefont {E.}~\bibnamefont {Lahoud}},
  \bibinfo {author} {\bibfnamefont {I.}~\bibnamefont {Shomroni}},\ and\
  \bibinfo {author} {\bibfnamefont {J.}~\bibnamefont {Steinhauer}},\ }\bibfield
   {title} {\bibinfo {title} {The a.c. and d.c. {J}osephson effects in a
  {B}ose--{E}instein condensate},\ }\href {https://doi.org/10.1038/nature06186}
  {\bibfield  {journal} {\bibinfo  {journal} {Nature}\ }\textbf {\bibinfo
  {volume} {449}},\ \bibinfo {pages} {579} (\bibinfo {year}
  {2007})}\BibitemShut {NoStop}%
\bibitem [{\citenamefont {Packard}\ and\ \citenamefont
  {Vitale}(1992)}]{Packard_1992}%
  \BibitemOpen
  \bibfield  {author} {\bibinfo {author} {\bibfnamefont {R.~E.}\ \bibnamefont
  {Packard}}\ and\ \bibinfo {author} {\bibfnamefont {S.}~\bibnamefont
  {Vitale}},\ }\bibfield  {title} {\bibinfo {title} {Principles of
  superfluid-helium gyroscopes},\ }\href
  {https://doi.org/10.1103/PhysRevB.46.3540} {\bibfield  {journal} {\bibinfo
  {journal} {Phys. Rev. B}\ }\textbf {\bibinfo {volume} {46}},\ \bibinfo
  {pages} {3540} (\bibinfo {year} {1992})}\BibitemShut {NoStop}%
\bibitem [{\citenamefont {Woffinden}\ \emph {et~al.}(2023)\citenamefont
  {Woffinden}, \citenamefont {Groszek}, \citenamefont {Gauthier}, \citenamefont
  {Mommers}, \citenamefont {Bromley}, \citenamefont {Haine}, \citenamefont
  {Rubinsztein-Dunlop}, \citenamefont {Davis}, \citenamefont {Neely},\ and\
  \citenamefont {Baker}}]{woffinden2022viability}%
  \BibitemOpen
  \bibfield  {author} {\bibinfo {author} {\bibfnamefont {C.~W.}\ \bibnamefont
  {Woffinden}}, \bibinfo {author} {\bibfnamefont {A.~J.}\ \bibnamefont
  {Groszek}}, \bibinfo {author} {\bibfnamefont {G.}~\bibnamefont {Gauthier}},
  \bibinfo {author} {\bibfnamefont {B.~J.}\ \bibnamefont {Mommers}}, \bibinfo
  {author} {\bibfnamefont {M.~W.~J.}\ \bibnamefont {Bromley}}, \bibinfo
  {author} {\bibfnamefont {S.~A.}\ \bibnamefont {Haine}}, \bibinfo {author}
  {\bibfnamefont {H.}~\bibnamefont {Rubinsztein-Dunlop}}, \bibinfo {author}
  {\bibfnamefont {M.~J.}\ \bibnamefont {Davis}}, \bibinfo {author}
  {\bibfnamefont {T.~W.}\ \bibnamefont {Neely}},\ and\ \bibinfo {author}
  {\bibfnamefont {M.}~\bibnamefont {Baker}},\ }\bibfield  {title} {\bibinfo
  {title} {{Viability of rotation sensing using phonon interferometry in
  {B}ose-{E}instein condensates}},\ }\href
  {https://doi.org/10.21468/SciPostPhys.15.4.128} {\bibfield  {journal}
  {\bibinfo  {journal} {SciPost Phys.}\ }\textbf {\bibinfo {volume} {15}},\
  \bibinfo {pages} {128} (\bibinfo {year} {2023})}\BibitemShut {NoStop}%
\bibitem [{\citenamefont {Gajdacz}\ \emph {et~al.}(2013)\citenamefont
  {Gajdacz}, \citenamefont {Pedersen}, \citenamefont {Mørch}, \citenamefont
  {Hilliard}, \citenamefont {Arlt},\ and\ \citenamefont
  {Sherson}}]{GajdaczMiroslav2013NFio}%
  \BibitemOpen
  \bibfield  {author} {\bibinfo {author} {\bibfnamefont {M.}~\bibnamefont
  {Gajdacz}}, \bibinfo {author} {\bibfnamefont {P.~L.}\ \bibnamefont
  {Pedersen}}, \bibinfo {author} {\bibfnamefont {T.}~\bibnamefont {Mørch}},
  \bibinfo {author} {\bibfnamefont {A.~J.}\ \bibnamefont {Hilliard}}, \bibinfo
  {author} {\bibfnamefont {J.}~\bibnamefont {Arlt}},\ and\ \bibinfo {author}
  {\bibfnamefont {J.~F.}\ \bibnamefont {Sherson}},\ }\bibfield  {title}
  {\bibinfo {title} {Non-destructive {F}araday imaging of dynamically
  controlled ultracold atoms},\ }\href@noop {} {\bibfield  {journal} {\bibinfo
  {journal} {Review of Scientific Instruments}\ }\textbf {\bibinfo {volume}
  {84}},\ \bibinfo {pages} {083105} (\bibinfo {year} {2013})}\BibitemShut
  {NoStop}%
\bibitem [{\citenamefont {Wilson}(2015)}]{Wilson_2015}%
  \BibitemOpen
  \bibfield  {author} {\bibinfo {author} {\bibfnamefont {K.~E.}\ \bibnamefont
  {Wilson}},\ }\emph {\bibinfo {title} {Developing a {T}oolkit for
  {E}xperimental {S}tudies of {T}wo-{D}imensional {Q}uantum {T}urbulence in
  {B}ose-{E}instein {C}ondensates}},\ \href
  {http://hdl.handle.net/10150/577309} {Ph.D. thesis},\ \bibinfo  {school} {The
  University of Arizona.} (\bibinfo {year} {2015})\BibitemShut {NoStop}%
\bibitem [{\citenamefont {Bauckhage}\ and\ \citenamefont
  {Sifa}(2015)}]{bauckhage2015k}%
  \BibitemOpen
  \bibfield  {author} {\bibinfo {author} {\bibfnamefont {C.}~\bibnamefont
  {Bauckhage}}\ and\ \bibinfo {author} {\bibfnamefont {R.}~\bibnamefont
  {Sifa}},\ }\bibfield  {title} {\bibinfo {title} {k-{M}axoids {C}lustering.},\
  }in\ \href {https://ceur-ws.org/Vol-1458} {\emph {\bibinfo {booktitle}
  {Proceedings of the {LWA} 2015 Workshops: KDML, FGWM, IR, and FGDB, Trier,
  Germany, October 7-9, 2015}}},\ \bibinfo {series} {{CEUR} Workshop
  Proceedings}, Vol.\ \bibinfo {volume} {1458},\ \bibinfo {editor} {edited by\
  \bibinfo {editor} {\bibfnamefont {R.}~\bibnamefont {Bergmann}}, \bibinfo
  {editor} {\bibfnamefont {S.}~\bibnamefont {G{\"{o}}rg}},\ and\ \bibinfo
  {editor} {\bibfnamefont {G.}~\bibnamefont {M{\"{u}}ller}}}\ (\bibinfo
  {publisher} {CEUR-WS.org},\ \bibinfo {year} {2015})\ pp.\ \bibinfo {pages}
  {133--144}\BibitemShut {NoStop}%
\bibitem [{\citenamefont {Rakonjac}\ \emph {et~al.}(2016)\citenamefont
  {Rakonjac}, \citenamefont {Marchant}, \citenamefont {Billam}, \citenamefont
  {Helm}, \citenamefont {Yu}, \citenamefont {Gardiner},\ and\ \citenamefont
  {Cornish}}]{Rakonjac2016}%
  \BibitemOpen
  \bibfield  {author} {\bibinfo {author} {\bibfnamefont {A.}~\bibnamefont
  {Rakonjac}}, \bibinfo {author} {\bibfnamefont {A.}~\bibnamefont {Marchant}},
  \bibinfo {author} {\bibfnamefont {T.}~\bibnamefont {Billam}}, \bibinfo
  {author} {\bibfnamefont {J.}~\bibnamefont {Helm}}, \bibinfo {author}
  {\bibfnamefont {M.}~\bibnamefont {Yu}}, \bibinfo {author} {\bibfnamefont
  {S.}~\bibnamefont {Gardiner}},\ and\ \bibinfo {author} {\bibfnamefont
  {S.}~\bibnamefont {Cornish}},\ }\bibfield  {title} {\bibinfo {title}
  {Measuring the disorder of vortex lattices in a {B}ose-{E}instein
  condensate},\ }\href@noop {} {\bibfield  {journal} {\bibinfo  {journal}
  {Phys. Rev. A}\ }\textbf {\bibinfo {volume} {93}},\ \bibinfo {pages} {013607}
  (\bibinfo {year} {2016})}\BibitemShut {NoStop}%
\bibitem [{\citenamefont {Gauthier}(2020)}]{Guillaume2020}%
  \BibitemOpen
  \bibfield  {author} {\bibinfo {author} {\bibfnamefont {G.}~\bibnamefont
  {Gauthier}},\ }\bibinfo {title} {Creation and {D}ynamics of {O}nsager
  {V}ortex {C}lusters},\ in\ \href
  {https://doi.org/10.1007/978-3-030-54967-1_6} {\emph {\bibinfo {booktitle}
  {Transport and {T}urbulence in {Q}uasi-{U}niform and {V}ersatile
  {B}ose-{E}instein {C}ondensates}}}\ (\bibinfo  {publisher} {Springer
  International Publishing},\ \bibinfo {year} {2020})\ pp.\ \bibinfo {pages}
  {139--169}\BibitemShut {NoStop}%
\bibitem [{\citenamefont {Gauthier}\ \emph {et~al.}(2019)\citenamefont
  {Gauthier}, \citenamefont {Reeves}, \citenamefont {Yu}, \citenamefont
  {Bradley}, \citenamefont {Baker}, \citenamefont {Bell}, \citenamefont
  {Rubinsztein-Dunlop}, \citenamefont {Davis},\ and\ \citenamefont
  {Neely}}]{Gauthier_2019}%
  \BibitemOpen
  \bibfield  {author} {\bibinfo {author} {\bibfnamefont {G.}~\bibnamefont
  {Gauthier}}, \bibinfo {author} {\bibfnamefont {M.~T.}\ \bibnamefont
  {Reeves}}, \bibinfo {author} {\bibfnamefont {X.}~\bibnamefont {Yu}}, \bibinfo
  {author} {\bibfnamefont {A.~S.}\ \bibnamefont {Bradley}}, \bibinfo {author}
  {\bibfnamefont {M.~A.}\ \bibnamefont {Baker}}, \bibinfo {author}
  {\bibfnamefont {T.~A.}\ \bibnamefont {Bell}}, \bibinfo {author}
  {\bibfnamefont {H.}~\bibnamefont {Rubinsztein-Dunlop}}, \bibinfo {author}
  {\bibfnamefont {M.~J.}\ \bibnamefont {Davis}},\ and\ \bibinfo {author}
  {\bibfnamefont {T.~W.}\ \bibnamefont {Neely}},\ }\bibfield  {title} {\bibinfo
  {title} {Giant vortex clusters in a two-dimensional quantum fluid},\
  }\href@noop {} {\bibfield  {journal} {\bibinfo  {journal} {Science}\ }\textbf
  {\bibinfo {volume} {364}},\ \bibinfo {pages} {1264} (\bibinfo {year}
  {2019})}\BibitemShut {NoStop}%
\end{thebibliography}%
\end{document}